\documentclass[manuscript]{aastex}
\usepackage{longtable}
\usepackage{natbib}

\usepackage{chngcntr}

\shorttitle{The HH 1165 jet}
\shortauthors{Riaz et al.}

\newcommand{\Msun}{M$_{\odot}$}

\usepackage{amsmath} % or simply amstext

\begin{document}

\title{First large scale Herbig-Haro jet driven by a proto-brown dwarf}

\author{B. Riaz}
\affil{Max-Planck-Institut f\"{u}r Extraterrestrische Physik, Giessenbachstrasse 1, 85748 Garching, Germany}

\author{C. Brice\~{n}o}
\affil{Cerro Tololo Inter-American Observatory, National Optical Astronomy Observatory, Casilla 603, La Serena, Chile}

\author{E. T. Whelan}
\affil{Maynooth University Department of Experimental Physics, National University of Ireland Maynooth, Maynooth Co. Kildare, Ireland}

\and

\author{S. Heathcote}
\affil{Cerro Tololo Inter-American Observatory, National Optical Astronomy Observatory, Casilla 603, La Serena, Chile}

\begin{abstract}

We report the discovery of a new Herbig-Haro jet, HH 1165, in SOAR narrow-band imaging of the vicinity of the $\sigma$ Orionis cluster. HH 1165 shows a spectacular extended and collimated spatial structure, with a projected length of 0.26 pc, a bent {\bf C}-shaped morphology, multiple knots, and fragmented bow-shocks at the apparent ends of the flow. The H$\alpha$ image shows a bright halo with a clumpy distribution of material seen around the driving source, and curved reflection nebulosity tracing the outflow cavities. The driving source of HH 1165 is a Class I proto-brown dwarf, Mayrit 1701117 (M1701117), with a total (dust+gas) mass of $\sim$36 $\rm M_{Jup}$  and a bolometric luminosity of $\sim$0.1 L$_{\odot}$. High-resolution VLT/UVES spectra of M1701117 show a wealth of emission lines indicative of strong outflow and accretion activity. 

SOAR/Goodman low-resolution spectra along the jet axis show an asymmetrical morphology for HH 1165. We find a puzzling picture wherein the north-west part exhibits a classical HH jet running into a pre-dominantly neutral medium, while the southern part resembles an externally irradiated jet. The {\bf C}-shaped bending in HH 1165 may be produced by the combined effects from the massive stars in the ionization front to the east, the $\sigma$ Orionis core to the west, and the close proximity to the B2-type star HR~1950. HH 1165 shows all of the signatures to be considered as a scaled-down version of parsec-length HH jets, and can be termed as the first sub-stellar analog of a protostellar HH jet system.

%We find a puzzling picture wherein the NW part of the outflow appears to be a classical HH jet running into a predominantly neutral medium, while the SE beam resembles an externally photo-ionized jet. 

%A comparison of the PV diagrams in the forbidden lines with the narrow-band images shows that the bright, blue-shifted emission that we see in the PV diagrams arises from the NW lobe of the jet, while the weak red-shifted wing is the faint lobe seen in the SE of the driving source 

\end{abstract}

\keywords{Herbig-Haro objects --- stars: jets --- Brown dwarfs --- stars: low-mass --- stars: winds, outflows --- open clusters and associations: individual ($\sigma$ Orionis)  }

\section{Introduction}

Herbig-Haro (HH) jets are so called as they are made up of a string of shocks or HH objects which represent individual ejection events (e.g., Reipurth \& Bally 2001). These atomic jets are most often associated with Class 0/I low mass young stellar objects (YSOs), and are important as they offer a way in which to study the mass loss history of the driving source (e.g., Frank et al. 2014). They can be parsec scale in length and their morphology is greatly affected by the ejection properties of the source and the environment in which they are propagating (e.g., Bally 2011). On scales of $<$ 0.1~pc, they are characterized by closely spaced knots which show evidence of internal shock waves likely caused by supersonic velocity jumps. It can be strongly argued that these velocity jumps are a result of intrinsic ejection variability (e.g., Bally et al. 2001; 2002; Frank et al. 2014). On parsec scales, the HH objects tend to be more sparsely distributed and fragmented. Interactions with the environment can also cause the HH jets to become bent, and to take on a {\bf C}- or {\bf S}-shaped morphology (e.g., Bally et al. 2006). 

While most of what is understood about outflows comes from observations of low mass YSOs, it is now 10 years since outflows were first associated with objects at the bottom of the mass spectrum of star forming regions i.e. young brown dwarfs (0.08~\Msun to 0.13~\Msun) (e.g., Whelan et al. 2005). Although numerous outflows have been detected in the mass range from brown dwarfs to very low mass stars (0.13~\Msun to 0.2~\Msun), in a majority of the cases, these are micro-jets with projected lengths of $\leq$0.03 pc (e.g., Whelan 2014). The brown dwarfs or very low-mass stars driving these micro-jets are all optically visible, and are therefore taken to be analogous to Class II YSOs. Some notable Class II brown dwarfs and very low-mass stars that are known to drive micro-jets are Par-Lup3-4 and ISO143, with projected lengths of $<$ 0.004~pc (Comer\'{o}n \& Fern\'{a}ndez 2011, Joergens et al. 2012, Whelan et al. 2012; 2014). Recently, we have identified a new micro-jet, HH~1158, which is driven by a Class Flat\footnote{The Class Flat is considered to be an intermediate stage between Class I and II (e.g., Greene et al. 1994).} very low-mass star, and has a projected length of $\sim$0.01 pc (Riaz \& Whelan 2015). 

Here we report the first observation of a large-scale ($\sim$0.26~pc) HH jet driven by a candidate proto-brown dwarf. This jet has been designated the HH number HH~1165. The driving source of the jet is a very low-luminosity Class I object, Mayrit~1701117 (hereafter M1701117), which lies at the cusp of the stellar/sub-stellar boundary and will likely evolve into a brown dwarf. HH~1165 shows many of the characteristics of a large-scale jet including multiple knots, bow-shocks, a {\bf C}-shaped morphology and a reflection nebula. The first hints of a jet driven by M1701117 were seen in moderate resolution spectra presented in Riaz et al. (2015). Follow-up UV-Visual Echelle Spectrometer (UVES) spectra provided more information on the jet close to the driving source. Images taken with SOAR have revealed the spectacular nature of the jet. In this paper, we present both the SOAR and UVES data. While the jet is not extended in the UVES data, much can still be learnt from the various accretion and outflow associated spectral line diagnostics. 

Sect.~\ref{obs} provides further information on the driving source, and a description of the various sets of observations and the data reduction. Sect.~\ref{results} presents the results from the data analysis of spectra and images. A discussion on the morphology of the jet and the different features observed is presented in Sect.~\ref{discussion}, along with a comparison with some of the well-known large-scale HH jets driven by protostellar objects.

\section{Target, Observations and Data Reduction} 
\label{obs}

\subsection{Target properties}
\label{source}

\begin{figure}[t]
\centering
\includegraphics[scale=0.57]{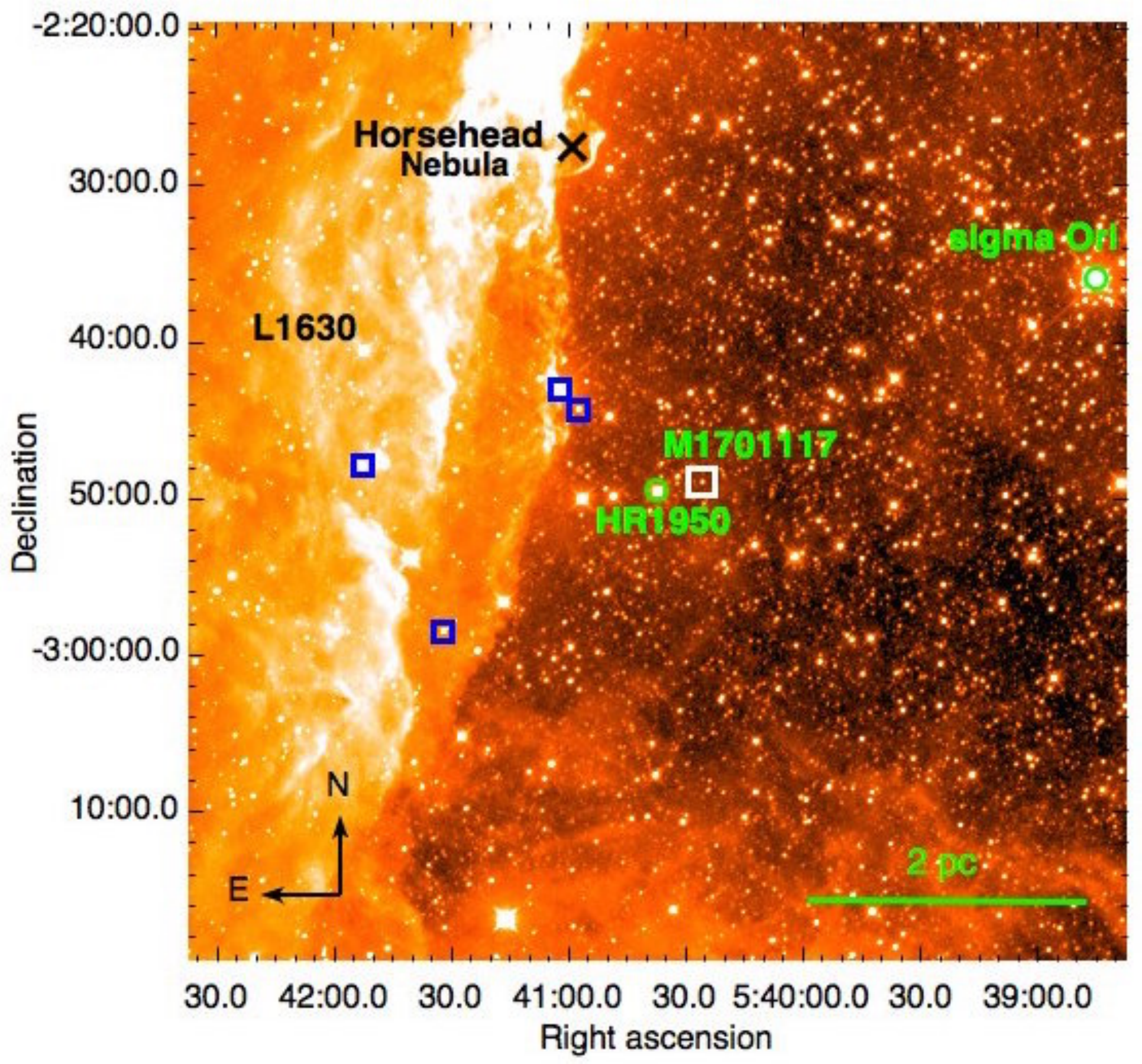}
\caption{The WISE 3.4$\mu$m 1x1 degree image showing the surroundings of the HH~1165 jet. Also marked are the locations of the $\sigma$ Orionis cluster core to the west, and the B2-type star HR~1950 located close ($\sim$0.33 pc) to the driving source M1701117. Blue squares mark the locations of some known B-type stars located near the ionization front seen brightly towards the east of M1701117.   }
\label{WISE}
\end{figure}

Figure~\ref{WISE} shows a wide 1$\times$1 degree view of the surroundings of the HH~1165 jet in the WISE mid-infrared 3.4$\mu$m image. The driving source M1701117 (05$^{h}$40$^{m}$25$^{s}$.8, -02$^{\circ}$48$^{\arcmin}$55.$^{\arcsec}$4) lies at the outer periphery of the $\sigma$~Orionis cluster (387$\pm$1.3 pc; Schaefer et al. 2016; Sim\'{o}n-Diaz et al. 2015). It is located $\sim$3.2 pc from the $\sigma$~Orionis cluster core, $\sim$1 pc from the south-west edge of the L1630 molecular cloud in Orion B, and also lies in close proximity to the massive B2IV-type star HR~1950, just $\sim$0.33 pc south-east from the source (Fig.~\ref{WISE}).

M1701117 was first discovered and catalogued by Caballero (2008). Riaz et al. (2015) investigated the multi-wavelength properties of this source and concluded that it is a very low-luminosity Class I object, with a total (gas + dust) mass of $\sim$36~M$_{Jup}$, and a bolometric luminosity of only $\sim$0.1 L$_{\sun}$. The mass of the central object in this Class I system can be constrained to within the sub-stellar mass regime ($\sim$0.04 -- 0.08 M$_{\sun}$), using the measured bolometric luminosity and numerical simulations of stellar evolution, as discussed in Riaz et al. (2016). Considering the very low mass reservoir in the (envelope+disk) for this system, and the fact that M1701117 drives a powerful outflow that will further dissipate the circumstellar envelope material, M1701117 will most likely evolve into a brown dwarf, and can be considered as a strong candidate proto-brown dwarf. Using multiple diagnostics observed in moderate resolution ($R\sim$1000) optical spectra, Riaz et al. (2015) estimated a mass accretion rate ($\dot{M}_{acc}$) of $\sim$ 6.4 $\times$ 10$^{-10}$~\Msun ~yr$^{-1}$ and a mass outflow rate ($\dot{M}_{out}$) of $\sim$ 1 $\times$ 10$^{-9}$~\Msun ~yr$^{-1}$, for M1701117.

%In this study we assume a mass and radius for M1701117 of 0.2~\Msun\ and 0.9~\Rsun\ and an extinction of A$_{v}$ = 2~mag. 

M1701117 is still considered a ``candidate'' member of $\sigma$ Orionis because of lying at the periphery of the cluster. The long Mayrit numbers imply that they are located far from the cluster centre, where there is contamination by other neighbouring young star-forming regions (Alnilam, $\epsilon$ Orionis, Horsehead Nebula). The latest measurement on the proper motion for M1701117 using 17 astrometric epochs from several surveys is $\mu_\alpha$ cos $\delta$, $\mu_\delta$ = -2$\pm$2, -6.3$\pm$1.9 mas/a (J. Caballero; {\it priv. comm.}). The proper motion of M1701117 is compatible within 1.0$\sigma$-2.5$\sigma$ with the mean value for the $\sigma$ Orionis cluster (+2.2$\pm$1, -0.5$\pm$1.0 mas/a), but also with that of other very young region towards Orion (except for the 32 Ori association). The third Gaia data release in 2018 can disentangle the tangential velocities of M1701117 and other young stars in the Horsehead Nebula and $\sigma$ Orionis area.

\subsection{High-resolution VLT/UVES spectroscopy}

High-resolution spectra of M1701117 were obtained with UVES (Dekker et al. 2000) in September, 2014. We used the standard DIC2 (437+760) setting, with cross dispersers of CD\#2 (HER\_5) and CD\#4 (BK7\_5). This setup provided a wavelength coverage from $\sim$3730 to 9460 {\AA} in one exposure. The slit width was set to 1$\arcsec$, resulting in a spectral resolution of $R \sim$ 40,000. We obtained two spectra for the target, one per slit orientation at a position angle (P.A.) of 0$\degr$ and 90$\degr$. The total on-source exposure time was 3200 seconds, split into two exposures. The seeing was recorded to be between 0.5$\arcsec$ and 1$\arcsec$ during the nights when these observations were made. The UVES spectra were reduced using the ESO Reflex pipeline. We estimate a signal-to-noise ratio (SNR) of better than $\sim$30 in the red arm and $\sim$10-15 in the blue arm of the combined spectrum.

\subsection{SOAR adaptive optics module observations}

Following on from the detection of outflow activity in M1701117 with UVES, narrow band optical imaging was obtained during two observing runs at the SOAR 4.1m telescope at Cerro Pach\'on, Chile. The first run was between Dec 13-16, 2015 and the second between Feb 29 - Mar 06, 2016. In order to obtain high angular resolution images, we used the SOAR Adaptive Optics Module (SAM) coupled with its imager SAMI. Technical details of the instrument can be found in (Tokovinin et al. 2010; 2012). SAMI provides a field-of-view (FOV) of $3\arcmin \times 3\arcmin$ with a pixel scale of $0.0455\arcsec$. All observations were obtained using a $2 \times 2$ binning, resulting in a $0.091\arcsec/pix$ scale. For these observations, we used the SOAR four inch square H$\alpha$ and [SII] narrow band filters, and the four inch square Bessell-R filter. The H$\alpha$ filter is centered at 6365 {\AA} and has a Full-Width at Half Maximum (FWHM) of 75 {\AA}. The [SII] filter central wavelength is 6738 {\AA} and has a FWHM=50 {\AA} \footnote{http://www.ctio.noao.edu/soar/content/filters-available-soar}. Table \ref{tabobs} gives an observing log indicating the dates, filters, and exposure times used.

\begin{figure}[t]
\center
\includegraphics[scale=0.7]{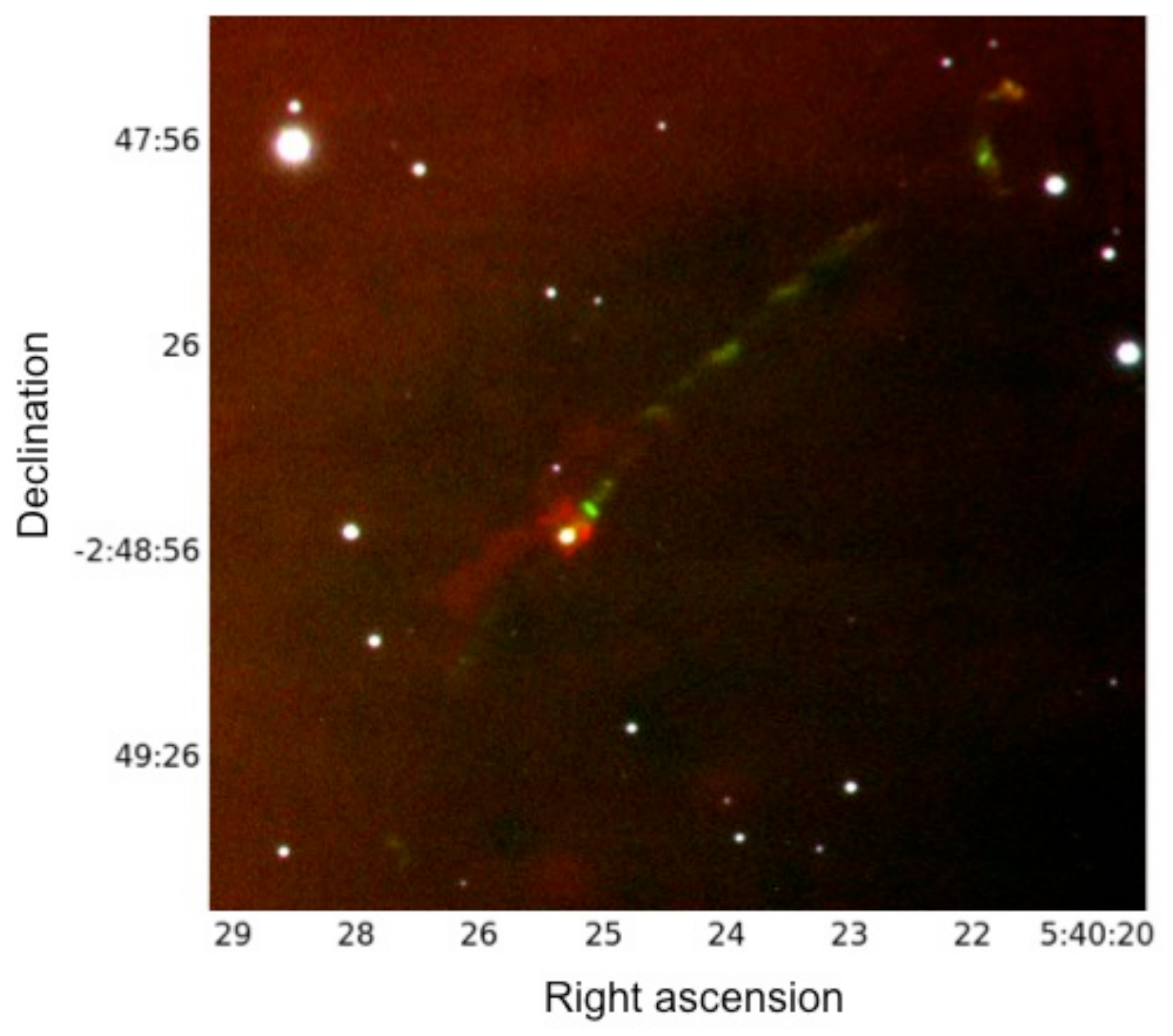}
\caption{The combined H$\alpha$ (red), [SII] (green), and $R$-band (blue) image for HH 1165. The image size is 3$\arcmin \times$3$\arcmin$ ($\sim0.3 \times$ 0.3 pc). North is up, east is to the left.  }
\label{comb-img}
\end{figure}

Each day we obtained a series of well exposed sky flats in each filter. The SOAR/SAMI data were reduced with the IRAF\footnote{IRAF is distributed by the National Optical Astronomy Observatory, which is operated by the Association of Universities for Research in Astronomy (AURA) under a cooperative agreement with the National Science Foundation.} {\sl ccdproc} and {\sl mscred} packages, which were used to do overscan and bias subtraction, flatfield corrections, and to merge the four image extensions produced by the four amplifiers that read each SAMI frame. For each filter, three processed images were finally median combined into the final science frame. An initial astrometric solution for each frame was obtained using the SAMI quick astrometry script {\sl samiqastrometry.py} \footnote{http://www.ctio.noao.edu/soar/content/reducing-your-sam-images}. Then, a refined astrometric solution was obtained using {\sl ccmap} in IRAF. During the SOAR SAM observations, sky conditions were not ideal, with seeing $\sim$1-1.5$\arcsec$, so we could not achieve the best possible image correction that the SAM Ground Layer AO system can deliver. The image FWHM of our combined science frames was in the range $\sim$0.5-0.7$\arcsec$. Figure~\ref{comb-img} shows the composite produced from the [SII], H$\alpha$, and $R$-band images for the HH~1165 jet; the individual images are shown in Fig.~\ref{Ha-SII-img}.

\subsection{SOAR Low Resolution Spectroscopy}

\begin{figure}[t]
\center
\includegraphics[scale=0.5]{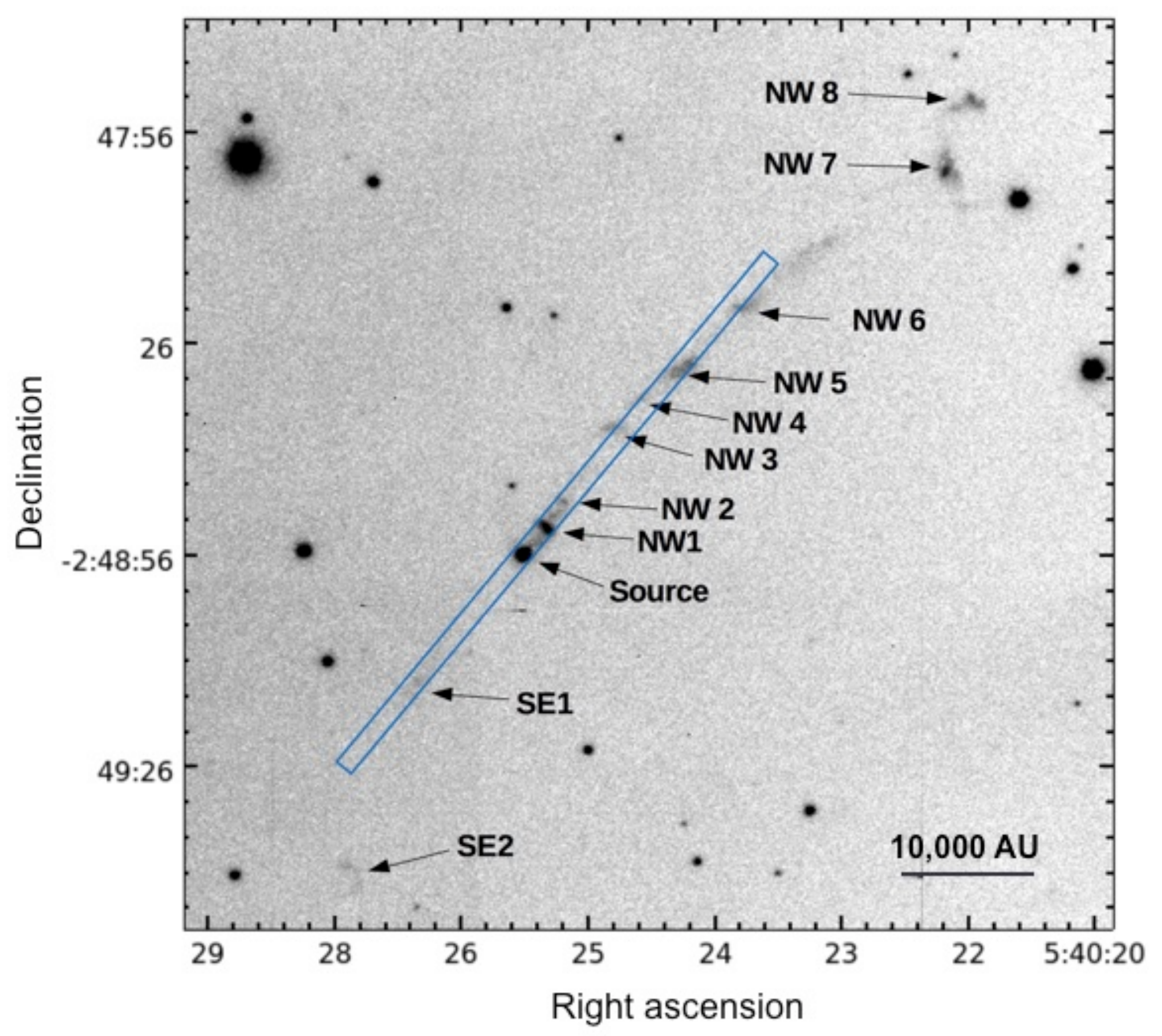}
\caption{The [SII]$\lambda$6731 image for HH 1165. Also labelled are the various knots observed in the NW and SE directions, as well as the slit orientation used for the Goodman spectra. The image size is 3$\arcmin \times$3$\arcmin$. North is up, east is to the left.  }
\label{SII-img}
\end{figure}

During the night of March 05, 2016, we obtained optical low resolution spectra at different positions along the HH~1165 jet using the Goodman High Throughput Spectrograph (GHTS) installed on the SOAR 4.1m telescope. The GTHS is a highly configurable imaging spectrograph that employs all transmissive optics and Volume Phase Holographic (VPH) Gratings. We used the 400 l/mm grating in its 400M1 preset mode, that provides a wavelength range $\sim$ 3100--7050 {\AA}. We kept the 0.15$\arcsec$ pix$^{-1}$ pixel scale in the dispersion direction, and binned the CCD by 2 in the spatial direction, providing a scale of 0.3$\arcsec$ per pixel. Combined with the $1.03\arcsec$ wide slit, this configuration produced a FWHM spectroscopic resolution of 6.3 {\AA}. Because time allowed only for obtaining a spectrum along a single orientation, we placed the slit at a position angle of $321.5\deg$ as shown in Fig.~\ref{SII-img}, a position that targets most of the features of the jet out to $\sim$50$\arcsec$. We obtained three 1800s long integrations, which were median combined after correcting for bias, and spatially registering the second and third exposure to the first one, which we used as reference. The basic image reduction was performed using standard IRAF {\sl ccdproc} and {\sl mscred} packages. The one-dimensional spectrum was extracted using routines in the IRAF {\it twodspec} and {\it onedspec} packages. For wavelength calibration we used a HgArNe lamp. The final reduced spectrum has a SNR$\sim$7 at 4000 {\AA} and $\sim$50 at H$\alpha$.

\section{Results}
\label{results}

An outflow for M1701117 was first detected in forbidden emission lines (FELs) observed in a low dispersion optical spectrum of this source (Riaz et al. 2015). While our UVES observations have further confirmed that M1701117 is indeed driving a powerful outflow, it is in the subsequent high angular resolution SOAR images that the impressive nature of the HH~1165 jet has become clear. Below, we firstly discuss what can be learnt about the morphology and structure of the jet from the SOAR images, and then outline what the UVES spectra reveal about the jet close to the driving source.

\subsection{Jet structure and morphology}

The wide-field images have revealed an arc minute long extended jet, with multiple shock structures. The various features in the jet are more clearly visible in the [SII] image, while the morphology close to the driving source shows some additional features that are only seen in the H$\alpha$ image.

\subsubsection{[SII] image}
\label{SII}

Figure~\ref{SII-img} shows the [SII] image for HH 1165. The jet extends to $\sim$1.5$^{\prime}$ ($\sim$0.17 pc) north-west (NW) of M1701117, with a PA$\sim$320$^{\circ}$. A counter-jet is seen extended to $\sim$0.8$^{\prime}$ ($\sim$0.09 pc) in the south-east (SE) direction from the source, with a PA$\sim$145$^{\circ}$. While at least 8 bright knots can be identified in the NW direction, there are just two knots seen in the SE part of the jet. The positions of the knots and their offset from the driving source are listed in Table~\ref{positions}. A very strong knot (NW1) is seen in the jet at 4.8$^{\prime\prime}$ NW from the source, with PA=316$^{\circ}$. The knot NW3 has a notable bow-shock structure. Of the knots farther to the NW, the strongest one (NW5) is located at $\sim$0.6$^{\prime}$ from the source, with PA$\sim$310$^{\circ}$. The knots NW7 and NW8 appear to be a fragmented bow-shock, at the apparent end of the NW jet. In particular, NW8 has a very different excitation from NW7 and the rest of the body of the jet (Fig.~\ref{SII-img}). We can interpret NW8 as (part of) the forward facing ``bow shock'' and NW7 as the Mach disk (reverse shock) in the working surface of the jet. Such excitation patterns are typical of the working surface in classical jets (e.g. HH~34; Reipurth et al. 2002). The morphology of NW7 and NW8 is particularly reminiscent of the the ``tip'' of the HH~33 jet (e.g., Mundt et al. 1984). Likewise, SE2 appears to be a fragmented, inverted bow-shock, and is likely the apparent end of the SE jet. The distance from the driving source to NW8 and SE2 are similar ($\sim$1$\arcmin$), suggesting a symmetric jet morphology. Note that since the apparent extent of the object fills our 3$\arcmin \times$3$\arcmin$ field of view, we cannot rule out the existence of more knots further from the source in either lobe. The spatial width of the brightest knots NW1, NW5, and NW7 is $\sim$1.7$\arcsec$ (0.003 pc), 4.2$\arcsec$ (0.007 pc), and 2.9$\arcsec$ (0.005~pc), respectively. The driving source itself appears most extended in the NW direction, with a width of 2.8$\arcsec$ (0.005~pc). The similarity in the widths of the knots both close to and far from the driving source indicates that this is a very collimated jet.

\begin{table}[h]
\centering
\caption{Positions of driving source and knots}
\begin{tabular}{cccc}
\hline
\hline
Object &   RA (J2000)  &   Dec (J2000) & Offset ($\arcsec$)  \\
\hline
M1701117   &    05:40:26.48 & -02:49:07.7 &   0.0$\pm$0.1  \\
   NW1     &  05:40:25.58 & -02:48:51.9 &      5.4$\pm$0.1   \\
   NW2      & 05:40:25.42 & -02:48:48.1 &      9.4$\pm$0.7  \\
   NW3      & 05:40:24.96 & -02:48:37.7 &     15.5$\pm$0.4  \\ 
   NW4      & 05:40:24.69 & -02:48:33.6  &     23.0$\pm$0.2  \\
   NW5      & 05:40:24.30 & -02:48:29.4  &     34.0$\pm$1.4   \\
   NW6    &  05:40:23.73 & -02:48:20.7 & 47.1$\pm$1.0 \\
   NW7  &  05:40:21.80 & -02:48:01.4 & 81.0$\pm$0.5 \\
   NW8  &  05:40:21.56 & -02:47:51.1 & 90.70$\pm$1.0 \\
   SE1     &  05:40:26.82 & -02:49:13.6   &   -24.9$\pm$0.9 \\
   SE2     &    05:40:27.47 & -02:49:40.1  & -50.8$\pm$2.0 \\
\hline            
\end{tabular}
\label{positions} 
\end{table}

A particular feature is the bent shape of the HH 1165 jet, such that the knots NW5-8 lie westward of the jet axis, and delineated by the inner knots (NW1 to NW3) and the driving source M1701117 (Fig.~\ref{SII-img}). The bending in the jet appears to begin at NW3-4. Looking at NW3 (Figs.~\ref{SII-img};~\ref{comb-img}), this knot has a partial bow shock structure facing North, and it has a nice trailing wing in H$\alpha$. It is reminiscent of some of the knots seen in the HH~34 and HH~111 jets (Reipurth et al. 2002; 1997). The NW3 knot could be the head of the outflow from a new outburst traveling along the NW1-3 axis, with the knots further out an older flow with a slightly different direction. A similar deflection is seen for the faint SE2 knot in the counter-jet (Fig.~\ref{SII-img}). HH~1165 shows the classical {\bf C}-shaped morphology often detected for jets in active regions. The possible causes for the observed bending in the jet are discussed in Section \ref{discussion}.

\subsubsection{H$\alpha$ image}
\label{halpha-img}

\begin{figure}[t]
\center
\includegraphics[scale=0.5]{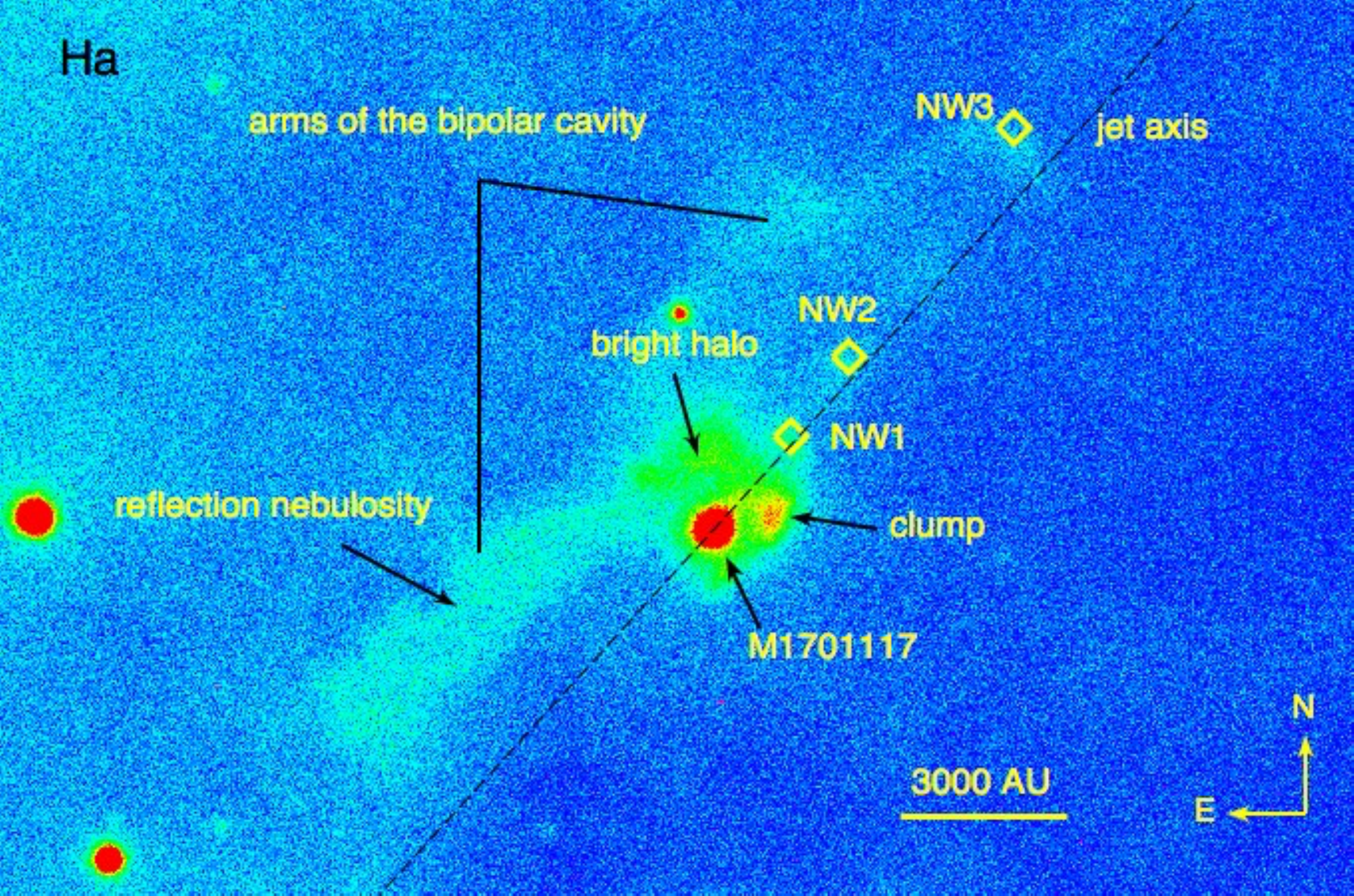} 
\caption{The H$\alpha$ image for HH 1165. We have roughly marked the bright reflection nebulosity tracing the outflow cavity on the eastern side of the source, the bright halo surrounding the driving source, and the bright clump seen within the halo. Solid line is the jet axis close to the driving source (PA$\sim$316$^{\circ}$). Also marked are the locations of the inner most NW1-3 knots as seen in the [SII] image. This is a zoomed-in 1$\arcmin \times$1$\arcmin$ image centered on M1701117. The linear scale is shown at the bottom. North is up, east is to the left.  }
\label{Ha-img}
\end{figure}

A spectacular feature in the H$\alpha$ image is the bright scattered emission seen close to the driving source (Fig.~\ref{Ha-img}). HH 1165 is associated with a curved reflection nebulosity on its eastern side. This extends to $\sim$20$\arcsec$ ($\sim$0.04 pc) to the SE and $\sim$14$\arcsec$ NW of the driving source. The fainter north-east nebulosity appears to be the mirror image of the bright south-east, elongated structure. There is, however, no evidence of reflection nebula on the western side of the source, which can be explained by the lower density and the lack of material to scatter on the south-west side (Sect.~\ref{photoevaporated}). Therefore, the bipolar nebulae can be described as one-sided. As a result it is difficult to correctly determine the symmetry axis of the reflection nebulosity, and whether or not it is coincident or offset from the outflow axis. Considering that the bright lobes on the eastern side appear to be extended along the outflow direction, the symmetry axis of the reflection nebulae is likely coincident with the outflow position angle. This suggests that the bright nebulosity traces the surface of dusty material in an outflow cavity. Outflow cavities are the polar regions of circumstellar envelope/disk systems that have been cleared of dense gas by stellar jets or wide-angle outflows (e.g., Li \& Shu 1996). The walls of outflow cavities are regions in which outflowing gas interacts with the infalling material in the envelope, producing scattered light emission, as seen in the H$\alpha$ images (e.g., Padgett et al. 1999). The presence of both a collimated jet and a broader wide-angle wind could carve out such wider, extended outflow cavities (Sect.~\ref{deflected}).

There is also evidence of a bright halo with a clumpy distribution of material seen around the driving source, which could be due to wind-envelope interaction. Some of the clumps are within the eastern outflow cavity, closer to the apparent apex of the cavity, which could be due to sub-structures in the cavity walls. A notably bright clump that is spatially distinct from the eastern cavities is seen at $\sim$2.7$^{\prime\prime}$ NW from the source, with PA$\sim$346$^{\circ}$ (Fig.~\ref{Ha-img}). This could be the inner, dense region of the western outflow cavity, and is perhaps a hint of the reflection nebulosity tracing the westward cavities. Such clump-like features are regularly produced at an offset from the main jet axis and close to the cavity walls in episodic jet and shock models, and are explained as the circumstellar (envelope) material being entrained by the jet/outflow (e.g., Machida 2014). An off-axis shock produced due to a mis-alignment between the disk and the outflow could also produce such clump-like features that are offset from the jet axis (Sect.~\ref{deflected}).

We note that the (H$\alpha$-$R$) continuum-subtracted image for HH~1165 (Fig.~\ref{R-Ha-img}; top panel) shows the same features of the reflection nebulosity, the bright halo and the clump, as seen in the H$\alpha$ image. This indicates that these features show excess emission in H$\alpha$, and are likely scattering the H$\alpha$ light either from the driving source M1701117 and/or the nearby ($\sim$0.3 pc) bright star HR~1950 (Fig.~\ref{WISE}). We have interpreted the extended SE tail seen in the H$\alpha$ image as ``reflection nebulosity''. This feature is likely produced by the interstellar dust reflecting the light from the nearby star HR~1950. A sign of this being reflection nebula is the presence of neutral He lines, and ionized C and Fe lines seen in the UVES spectra (Fig.~\ref{fullspec}), which could be the contribution from HR~1950 that acts as the illuminating star. If this were an emission nebula, we would see strong emission from ionized gas. Therefore, we hypothesize that this feature is due to scattered light from external material. The bright clump and the halo features are seen in scattered light emission from the circumstellar material internal to the Class I source M1701117. The clump is seen within the halo, and the size of the halo is comparable to the size of the envelope in M1701117 as estimated from SED modeling ($\sim$1500 AU; Riaz et al. 2015). As this clump is dominated by H$\alpha$ excess emission in the continuum-subtracted H$\alpha$ image (Fig.~\ref{R-Ha-img}), the likely cause of the ionization is a hard shock. Future polarization studies can help understand the origin of the scattered light emission, while spectroscopy along the extended SE tail can reveal if this is reflection or emission nebula.

%, both of which show strong emission in H$\alpha$

The bottom panel in Fig.~\ref{R-Ha-img} shows the (H$\alpha$-[SII]) image for HH~1165. An interesting new feature revealed in this image is a negative (low H$\alpha$ flux, high [SII] flux) patch located very close ($\sim$0.001 pc) to the driving source (labelled as ``unresolved knot''). This could be an unresolved shock emission knot that is bright in [SII], and produces the prominent shock emission lines seen in the UVES spectra for M1701117 (Fig.~\ref{fullspec}). 
Another notable feature is the closest knot NW1, which is also negative in the (H$\alpha$-[SII]) image, and is indeed the brightest knot seen in the [SII] image. None of the other NW knots are visible in (H$\alpha$-[SII]), indicating similar emission in H$\alpha$ and [SII]. The other features seen in the H$\alpha$ image are also brightly seen here, consistent with our earlier argument that these features show excess flux in H$\alpha$.

%The other notable thing is the farthest NW7 knot which is also negative in Ha-SII, but none of the other NW knots are visible here, even though we see traces of them in the Ha image.

%The extended SE feature could also be a ``cometary proplyd tail'' (Sect.~\ref{photoevaporated}), analogous to the tails seen for the externally irradiated HH 444-445 jets in $\sigma$ Orionis (Reipurth et al. 1998). 

\subsection{UVES spectroscopy}

\begin{figure}
\center
\includegraphics[scale=0.5]{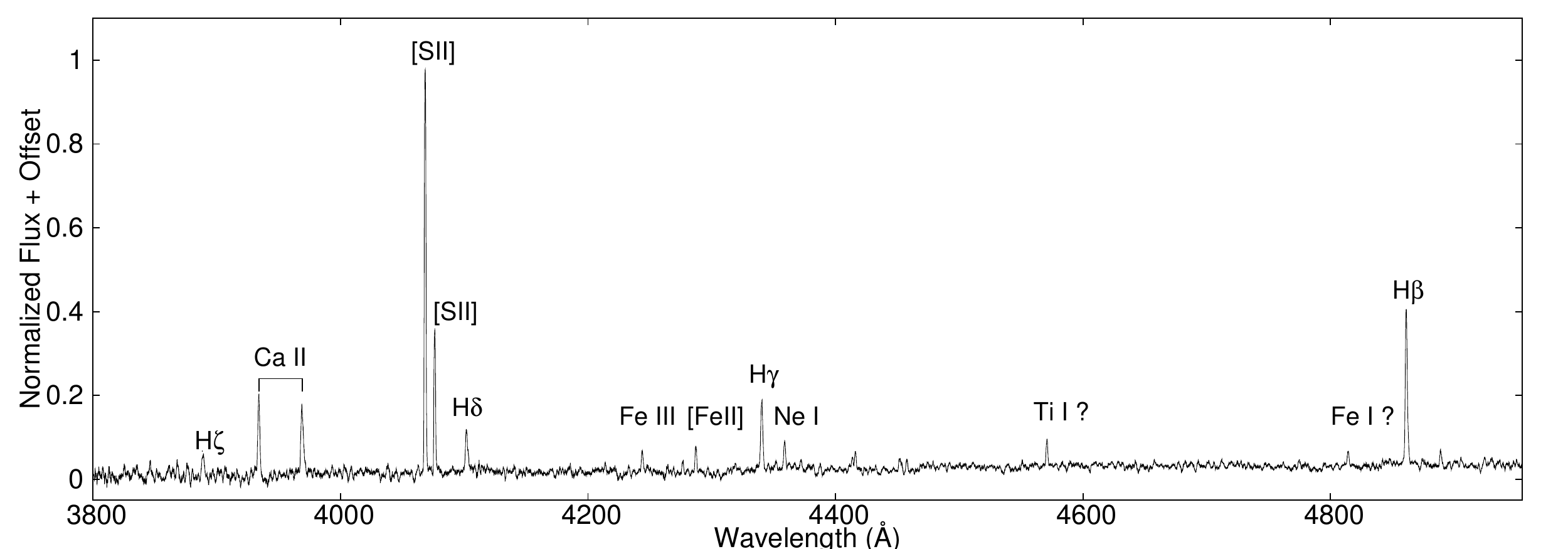}
\includegraphics[scale=0.5]{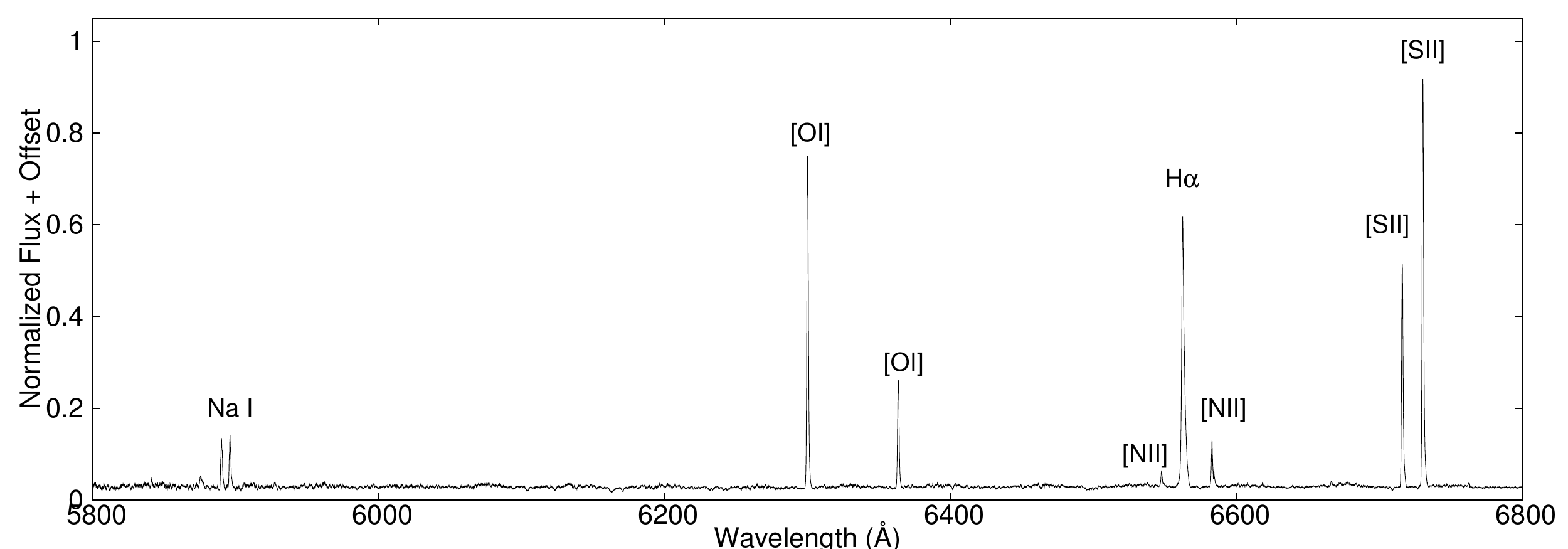}
\includegraphics[scale=0.5]{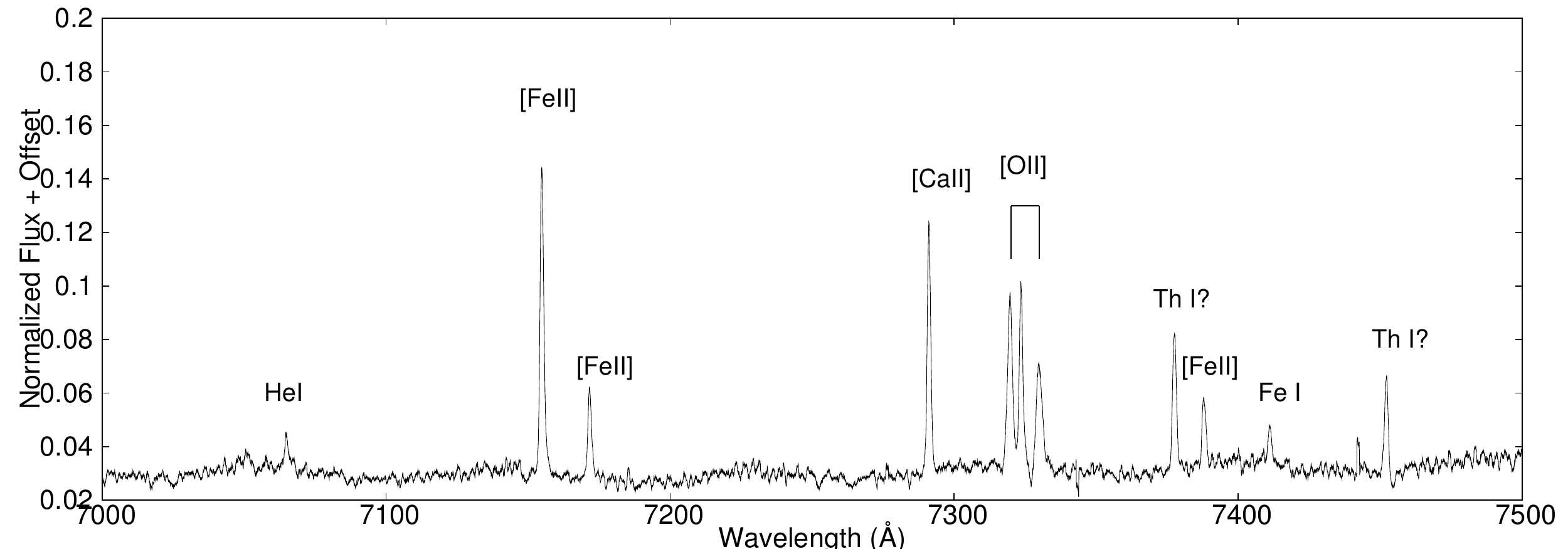}
\includegraphics[scale=0.5]{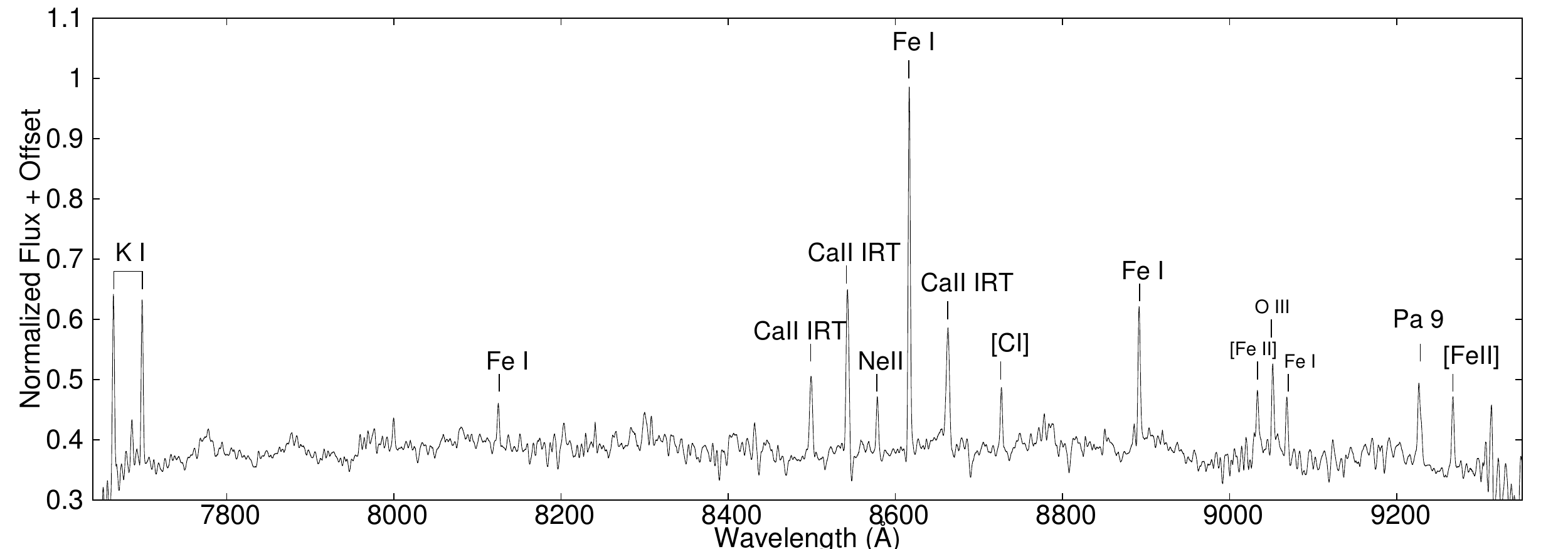}
\caption{UVES spectra for M1701117 with the prominent accretion- and outflow-associated emission lines marked. }
\label{fullspec}
\end{figure}

The UVES spectra show a wealth of emission lines indicative of strong outflow and accretion activity. The full wavelength range in the UVES spectra are shown in Figure \ref{fullspec}, and the fluxes and equivalent widths of the identified lines are given in Table \ref{fluxes}. On the blue side of the spectrum, the Balmer lines of H$\alpha$ and H$\beta$ are strongly detected, both of which are notable accretion tracers, while there is a weak detection for the H$\gamma$, H$\delta$ and H$\zeta$ lines. Note that the width of the H$\alpha$ profile at 10\% of the line peak is 280 km s$^{-1}$, larger than the $\geq$200 km s$^{-1}$ typically considered as a threshold to distinguish between accreting and non-accreting very low-mass/substellar sources (e.g., Muzerolle et al. 2003). There is also prominent emission in the Ca II H-line at 3968.5 {\AA} and the K line at 3933.7 {\AA}, while the Na I doublet at 5893 {\AA} is seen in emission, another indication of strong accretion activity. 

Among the outflow associated lines, there is strong emission seen in the [SII]$\lambda\lambda$4068,4076, [OI]$\lambda$$\lambda$6300, 6363, [SII]$\lambda$$\lambda$6716, 6731, and the [NII]$\lambda$6583 FELs (Fig.~\ref{fullspec}). The very strong emission in the [SII] lines, with [SII](6716+6731)/H$\alpha$ $>$ 1 (Sect.~\ref{ratios}) is typical of the low velocity shocks in classical HH jets, for e.g., the spectrum at the base of the HH~35 jet close to its Class I protostellar source shows similar sulphur enhanced emission (e.g. Ogura et al. 1995). There is also a weak detection in the [NII]$\lambda$6548 line, which is clearly resolved from the H$\alpha$ line. The presence of this rarely seen FEL was also noted in the spectrum of a jet driven by another very low-mass/sub-stellar object Mayrit 1082188 (Riaz \& Whelan 2015). The UVES spectra have reinforced that HH~1165 exhibits shock emission lines commonly seen in classic HH jets, and the low excitation lines of [SII] [OI] and [NII] being relatively stronger compared to H$\alpha$ suggests that the jet is propagating into a pre-dominantly neutral medium, as further discussed in Sect.~\ref{ratios}. 

%The [NII]$\lambda$6547 line requires a higher temperature and density than the [NII]$\lambda$6583 line, and so is not commonly detected in very low-mass stars. However, the low density limit for the [NII](6548/6583) line ratio ($\sim$0.3) is typical for a classic HH jet driven by a protostar, but the first such case for a proto-brown dwarf. 

In the red part of the spectrum (Fig.~\ref{fullspec}), there is strong emission in the outflow FELs of [FeII] at 7155 {\AA}, 7172 {\AA}, and 7388 {\AA}. Also notable is emission in the Ca II infrared triplet and the Paschen 9 line, typically associated with strong accretion. We see the K I$\lambda\lambda$7665,7699 doublet in emission, which is a well-known signature of the presence of an outflow wind (e.g., Hillenbrand et al. 2012). The 7320 {\AA} - 7330 {\AA} [OII] doublet is detected in emission, which is suggested to be kinematically associated with a strong outflow, and possibly has a formation in high-velocity winds (e.g., Hillenbrand et al. 2012). The HH 1165 spectrum also exhibits several other atomic lines of He I, [Ca II], Fe I, Th I, Ne II, [C I], Ti I, Ne I, O III (Figs.~\ref{fullspec}), most of which are seen in emission and are characteristic of young accreting stars (e.g., Fischer et al. 2008). The Fe I line at 8616 {\AA}, in particular, is very strong. We have also noted emission in the low-excitation line of Ti I, and the high-excitation line of C I, which have been observed earlier in strong outflow sources (e.g. Covey et al. 2011).

The aim of the high spectral resolution UVES project was to conduct a spectro-astrometric analysis of the outflow tracers (Sect.~\ref{SAanalysis}), and to obtain robust measurements on $\dot{M}_{acc}$ and $\dot{M}_{out}$ using multiple line diagnostics (Sect.~\ref{activity}).

\subsubsection{Spectro-astrometric analysis}
\label{SAanalysis}

\begin{figure}
\centering
\includegraphics[width=15cm]{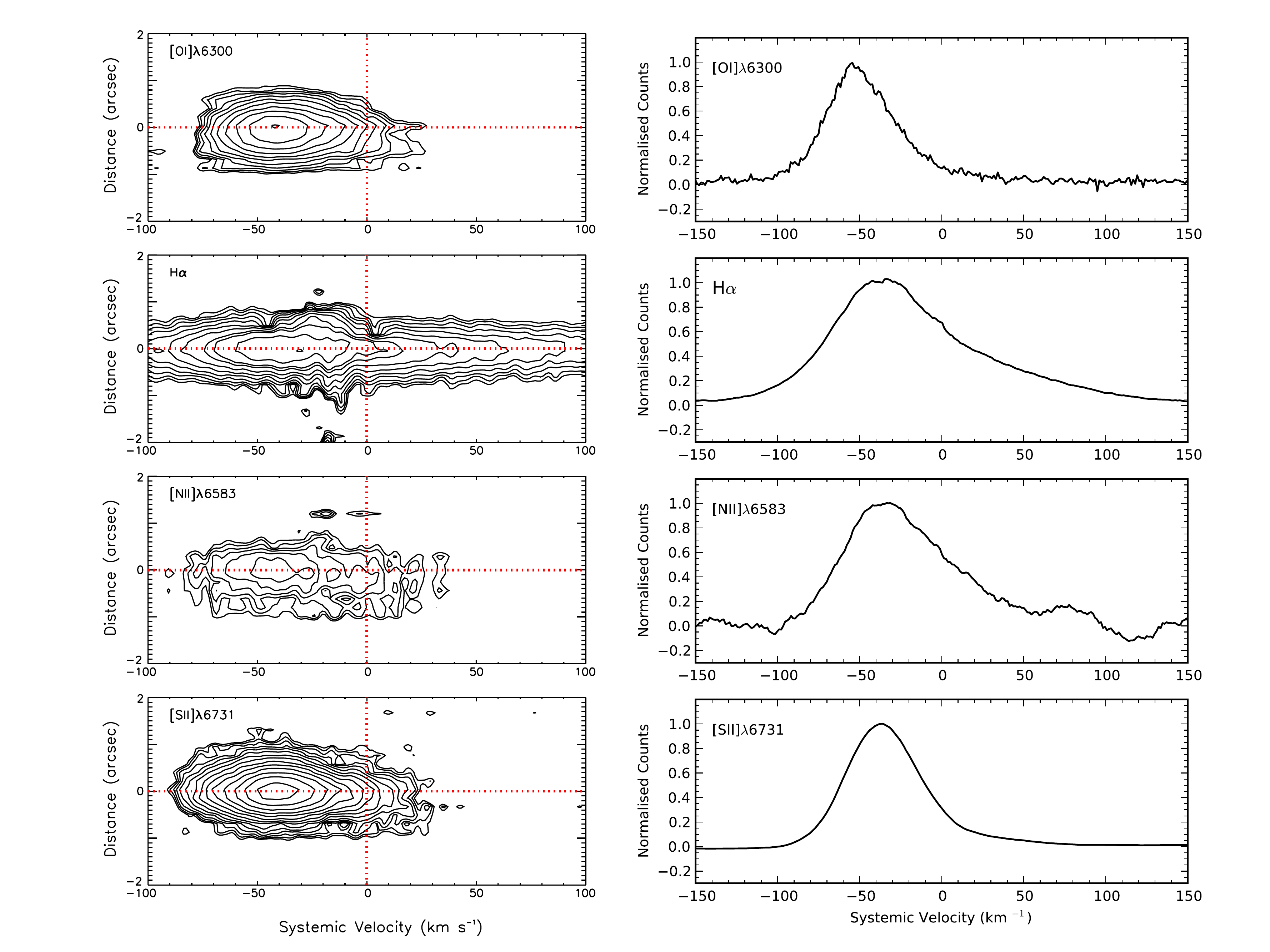}
\caption{Position-velocity diagrams of the FELs and H$\alpha$ line in the 0$^{\circ}$ spectrum of M1701117. The continuum has been subtracted. The line regions have been Gaussian smoothed to increase the SNR. Contours begin at 3-$\sigma$ and increase with a log scale. A clear blue-shifted knot is seen and a red-shifted wing. The emission is not spatially extended.}
\label{pvd}
\end{figure}

Figure~\ref{pvd} shows the position-velocity (PV) diagrams (left) and the line profiles (right) for the [OI]$\lambda$6300, [NII]$\lambda$6583, [SII]$\lambda$6731 FELs, and the H$\alpha$ line at the 0$^{\circ}$ slit PA. Similar results are seen for the [NII]$\lambda$6548 and [SII]$\lambda$6716 FELs at the 90$^{\circ}$ slit PA. All emission lines show a broad profile, with line widths of $>$50 km s$^{-1}$. Bright sky lines ([OI]) and nebular lines (H$\alpha$, [NII], [SII]) had to be removed and we can see evidence of the subtraction at $\sim$ 0 km s$^{-1}$ position (e.g. H$\alpha$ line). The velocities shown are systemic; we have considered the mean (30$\pm$10 km s$^{-1}$) of the radial velocity measurements for several members in the $\sigma$ Orionis cluster (e.g., Sacco et al. 2009). HH 1165 shows an asymmetric jet morphology, where the peak in emission is blue-shifted to $\sim$ 45 km s$^{-1}$ for all FELs (Fig.~\ref{pvd}). There is also a faint red-shifted wing, which is more apparent in the [NII]$\lambda$6583 line profile than in the PV plots. A comparison of the PV diagrams in the forbidden lines with the narrow-band images shows that the bright, blue-shifted emission that we see in the PV diagrams arises from the NW lobe of the jet, while the weak red-shifted wing is the faint lobe seen in the SE of the driving source (Figs.~\ref{pvd};~\ref{SII-img}).

%Since none of the emission lines are spatially resolved, we have conducted a spectro-astrometric analysis to extract the spatial information. 

The two components, however, are unresolved and a spectro-astrometric analysis is required to extract the spatial information. Figure~\ref{spectro} shows the spectro-astrometric analysis of the H$\alpha$, [NII]$\lambda$6583 and [SII]$\lambda$6731 lines, for both the 0$^{\circ}$ (black) and the 90$^{\circ}$ (red) slit PA spectra. The spectro-astrometric analysis is for the spectra without continuum subtraction, in order to enhance any offsets in the weak red-shifted wing. The offsets in the [OI]$\lambda$6300 lines are at or below the 1-$\sigma$ noise level of the measurement. This is due to the larger critical density of the [OI]$\lambda$ 6300 line, which traces emission much closer to the source (e.g., Whelan et al. 2009ab; Whelan et al. 2014). Bipolar offsets are seen in all FELs as well as in the H$\alpha$ line, with the blue-shifted peak and red-shifted wing offset in opposite directions. One interesting result here is that the bulk of the H$\alpha$ emission is offset in the outflow. Also note that the shape and width of the H$\alpha$ line is the same as the FELs, suggesting that the H$\alpha$ emission is mainly dominated by the outflow emission. We would not expect such a significant offset if the H$\alpha$ line was primarily tracing accretion. It may be the case that the H$\alpha$ emission from the accretion zone is obscured by the envelope/disk material surrounding the central source.

\begin{figure}[h]
\centering
\includegraphics[width=4.8cm]{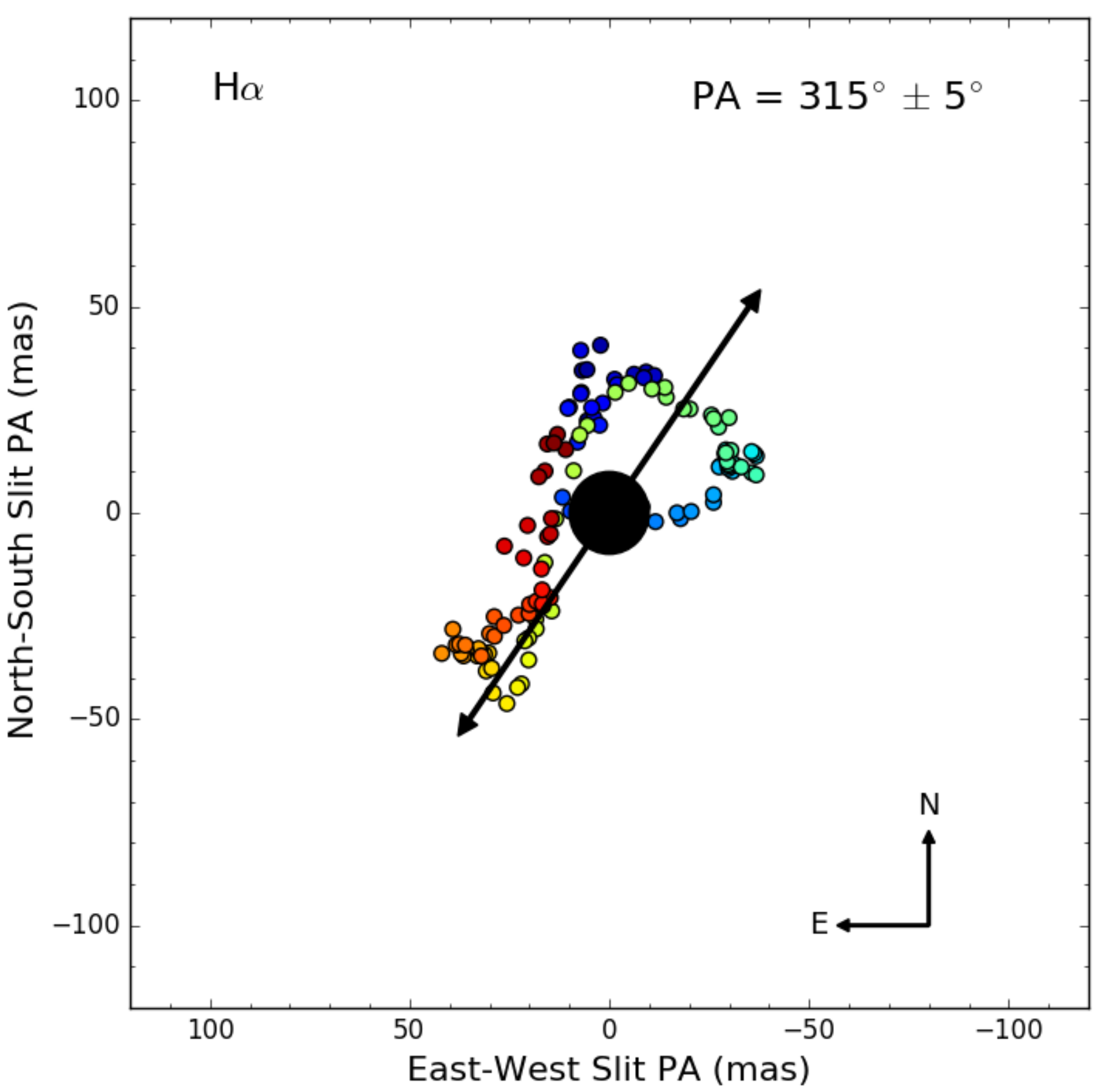}
\includegraphics[width=4.8cm]{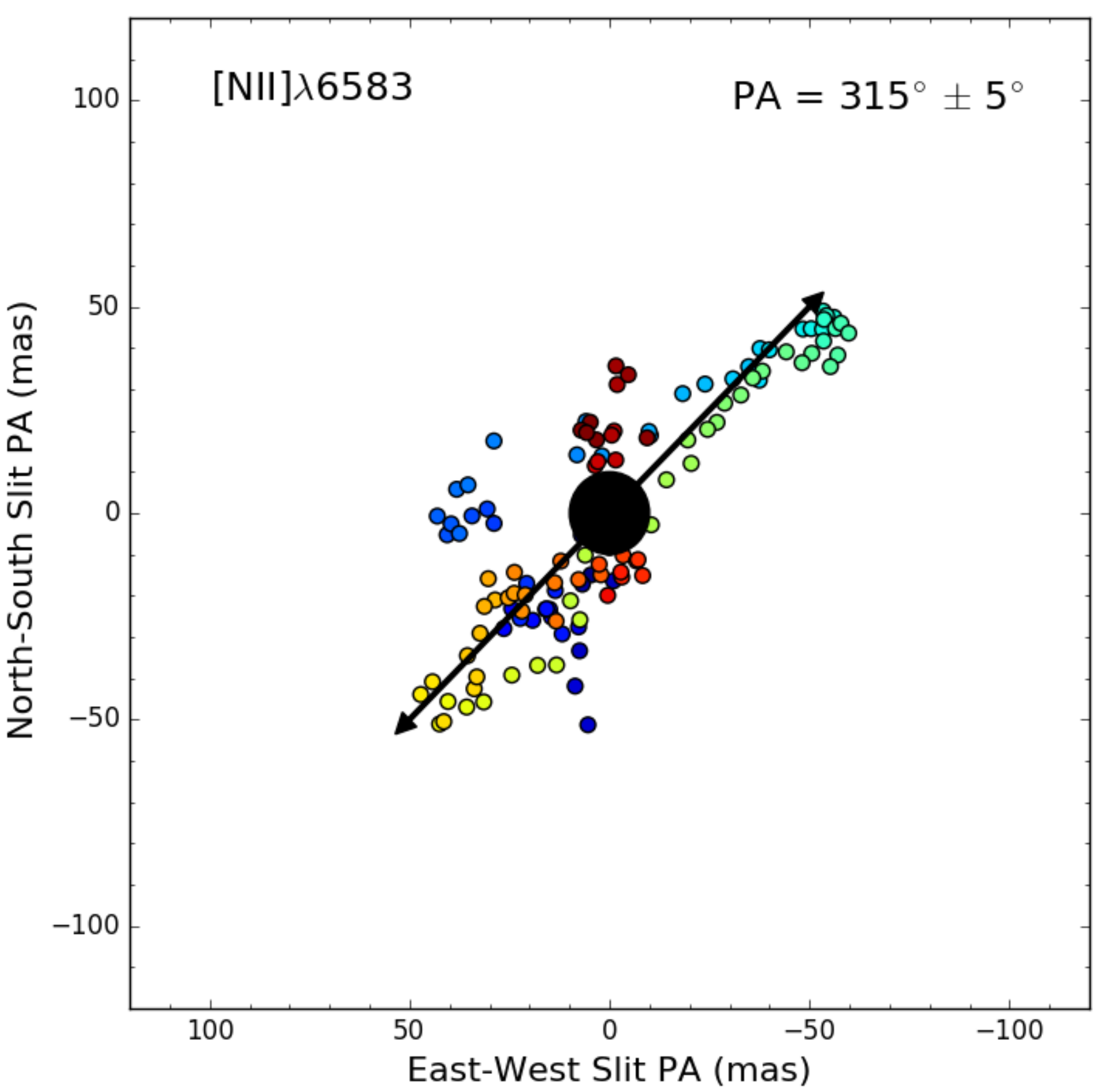}
\includegraphics[width=5.8cm]{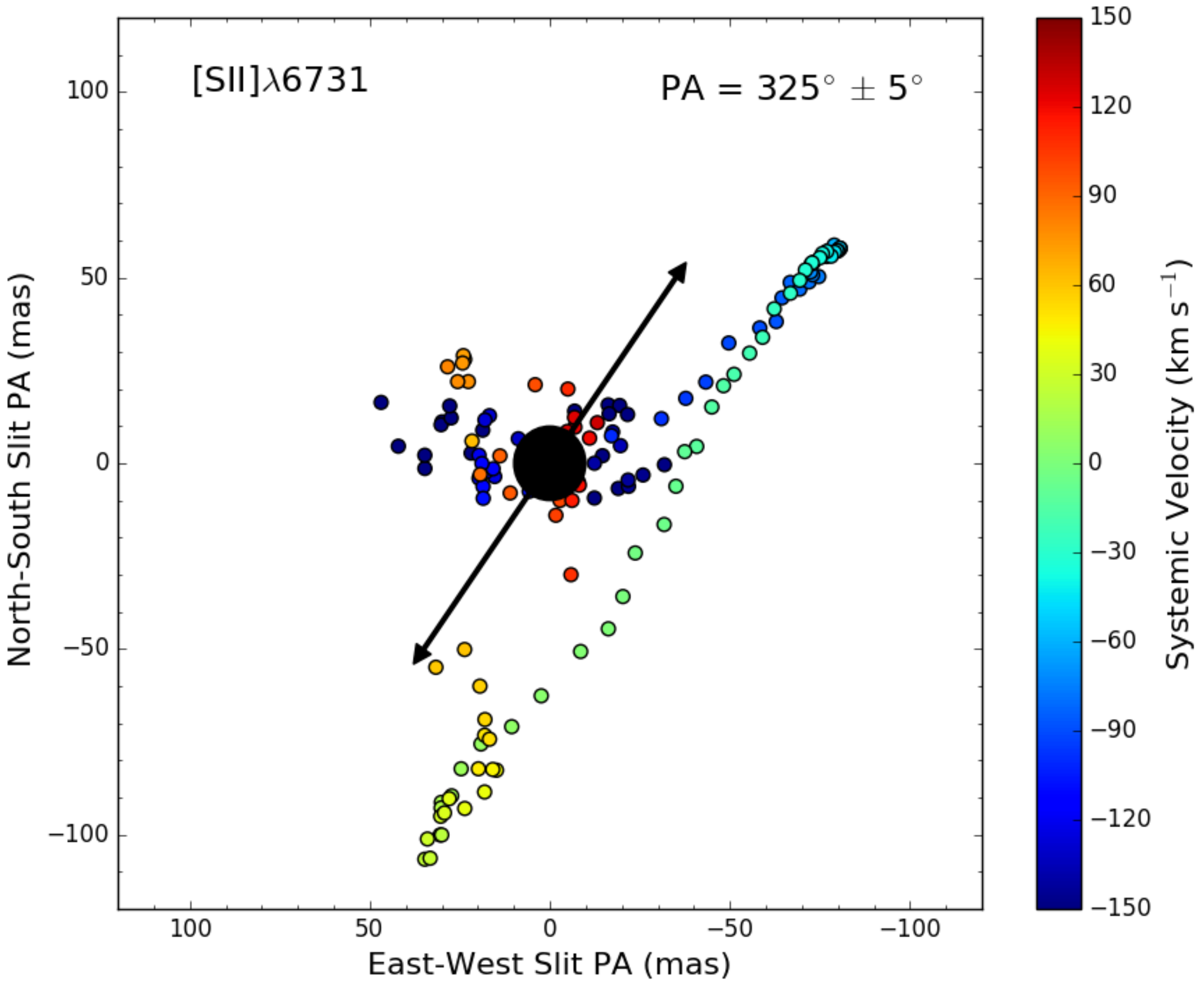}
\caption{Combining the spectro-astrometric results from the 0$^{\circ}$ and 90$^{\circ}$ to study the 2D spatial information. We fit the extended emission to estimate the jet PA. The filled black circle represents the uncertainty in the spectro-astrometric analysis of the continuum. The emission at 0 km s$^{-1}$ is offset from the continuum position in the [NII] and [SII] lines, which is a scattering effect. }
\label{jetPA}
\end{figure}

In Figure \ref{jetPA} we have combined the offsets at the slit PAs of 0$\degr$ and 90$\degr$ to recover the PA of the outflow on the sky. By fitting the PA of the extended emission, we get an average PA of 318$\degr$$\pm$5$\degr$, which is in strong agreement with the jet PA delineated by the inner knots in the [SII] image. This is to be expected as SA has been shown to be a very powerful tool for recovering jet PAs in the case where the jet emission is spatially unresolved (Whelan et al. 2012). A further notable result is that the emission at 0 km s$^{-1}$ is offset from the continuum position. This is especially clear in the case of the [NII]$\lambda$6583 and [SII]$\lambda$6731 lines in Figure \ref{spectro}. We would expect the jet emission to be coincident with the source position at 0 km s$^{-1}$. The fact that this effect is not as apparent in the H$\alpha$ line as compared to the FELs suggests that this is a scattering effect, and we are not seeing the jet emission at the driving source but at some small distance from the source.

\subsubsection{Accretion and outflow activity rates}
\label{activity}

The accretion rates for M1701117 have been measured using the luminosity of 13 different accretion tracers (taken from the UVES data), and an $A_{V}$$\sim$2 mag. We have excluded the lines that were blended or too noisy. The line luminosities measured for these tracers were extinction-corrected and then converted into the accretion luminosity and the accretion rate $\dot{M}_{acc}$ following the line luminosity relationships from Alc\'{a}la et al. (2014). The $\dot{M}_{acc}$ measurements are listed in Table~\ref{acc-rates} and the results are presented in Fig.~\ref{accretion}. The average $\dot{M}_{acc}$ was measured at (9 $\pm$ 7) $\times$10$^{-10}$ $M_{\sun}$ yr$^{-1}$. This is in agreement with the value of  $\dot{M}_{acc}$ = 6.4 $\times$10$^{-10}$ $M_{\sun}$ yr$^{-1}$ published in Riaz et al. (2015). The analysis also includes the indirect tracer [OI]$\lambda$6300, which is assumed to originate from above the obscuration zone of the circumstellar material. We have not included the [OI]$\lambda$6300 line in calculating the average value of $\dot{M}_{acc}$. Note that the difference between $\dot{M}_{acc}$ as estimated from the [OI]$\lambda$6300 line (18 $\times$10$^{-10}$ $M_{\sun}$ yr$^{-1}$) and the average $\dot{M}_{acc}$ is not as significant here as for the values reported in Riaz et al. (2015). It is argued that in the cases where the accretion region is obscured, the [OI]$\lambda$6300 line may give a more accurate estimate of the accretion rate (e.g., Herczeg \& Hillenbrand 2008).

%Using the results of \cite{Alcala14}, the line luminosity (L$_{line}$) could be converted to the accretion luminosity (L$_{acc}$) and then the equation $\dot{M}_{acc}$ = 1.25 (L$_{acc}$ R$_{*}$) / (G M$_{*}$) used \citep{Gullbring98, Hartmann98} to find $\dot{M}_{acc}$. R$_{*}$and M$_{*}$ were taken as 0.9~\Rsun\ and 0.2~\Msun\ respectively.

%We obtained estimates on the disc mass accretion rate, M? acc , using multiple diagnostics of the Balmer and Ca ii IRT lines (Table 3), and the latest line luminosity relationships from Alcala? et al. (2014). 

One important consideration is that the accretion tracers may not sample well the accretion zone due to the obscuring and scattering effects of the envelope/disk around the source. Indeed, the spectro-astrometric analysis indicates that a large part of the H$\alpha$ line is actually coming from the jet and not the accretion zone (Fig.~\ref{spectro}). As also noted in Fig.~\ref{spectro}, there are similarities in the H$\alpha$ emission line profile and those of the FELs. Therefore, H$\alpha$ may not provide a robust estimate of $\dot{M}_{acc}$. To investigate this further, we have compared in Fig.~\ref{accretion_lines} the shape of all accretion line diagnostics used to calculate $\dot{M}_{acc}$ with the shape of the outflow-associated [SII]$\lambda$ 6731 line. The first thing to note is that the [OI]$\lambda$6300 and [SII]$\lambda$ 6731 lines are almost a perfect match, as would be expected. The Balmer lines closely resemble the FELs, such that their peaks match the peak of the [SII]$\lambda$ 6731 line, and they all have a red-shifted wing. The He I 7065 {\AA} line matches almost perfectly to the [SII]$\lambda$ 6731 line. The Ca II lines are very different from the FELs in that they are much broader and the line peaks do not match. We also find the Balmer lines to be comparatively broader than the [SII] line, with the broader component likely being the underlying accretion component. The comparison with the Na I line is similar with the line being double peaked due to the interstellar absorption component.

Thus, there is likely a significant outflow component in the Balmer, Na and He lines that is not seen in the Ca II lines. We are seeing a suppressed accretion component in the Ca II lines, while the Balmer line profiles are made up of a suppressed accretion component plus an outflow component. Note that it was not possible to check for an extended outflow component in the He I and Na I lines using a spectro-astrometric analysis, due to the weakness of the line (He I) or the continuum emission in the vicinity of the line (Na I). Overall our analysis shows that it is difficult to estimate $\dot{M}_{acc}$ from the line luminosities due to the obscuration of the accretion region and the significant contribution from outflow. These factors explain the spread in the values plotted in Figure \ref{accretion} and the resultant large error on $\dot{M}_{acc}$.

The mass outflow rate, $\dot{M}_{out}$, for M1701117 has been derived from the luminosity of the [SII]$\lambda$6731 line, using the UVES spectra and following the method of Hartigan et al. (1995). This method requires an estimate for $n_{e}$, the critical density of [SII] ($n_{cr}$), the size of the aperture on the plane of the sky, and the tangential velocity. We used $n_{e}$ = 5000 cm$^{-3}$ (as measured at the source position from the UVES spectra), $n_{cr}$ = 1.3 x 10$^{4}$ cm$^{-3}$, V$_{tan}$ = 80 km s$^{-1}$ and the size of the aperture (l$_{tan}$) = 1$\arcsec$ (the slit width). The tangential velocity of (80$\pm$10)~km s$^{-1}$ is determined from the peak velocity of the FELs (30$\pm$5)~km s$^{-1}$ and the inclination angle (50$^{\circ}$ -- 65$^{\circ}$) for the system (Riaz et al. 2015).

We estimate $\dot{M}_{out}$ of (35 $\pm$ 17)$\times$10$^{-10}$ $M_{\sun}$ yr$^{-1}$ for M1701117. This is an estimate of $\dot{M}_{out}$ in the part of the outflow closest to the driving source. The estimate on $\dot{M}_{out}$ presented here is higher than that given in Riaz et al. (2015). This is due to the larger value of the tangential velocity used here. The high spectral resolution of UVES means that the radial velocities of the FELs can be well measured. Uncertainties which persist are related to the effect of obscuration and scattering on the accretion tracers and the extinction towards the jet. Therefore, we draw the same conclusions as in Riaz et al. (2015) that this analysis needs to be improved. Nevertheless, we note that the $\dot{M}_{out}$ and $\dot{M}_{acc}$ measured for M1701117 are an order of magnitude higher than the mean rates of the order of 10$^{-10}$ $M_{\sun}$ yr$^{-1}$ estimated for Class II very low-mass stars and brown dwarfs (e.g., Whelan et al. 2009ab), which is consistent with the earlier Class I evolutionary stage for M1701117.

\subsection{Variations in line intensity and density along the jet}
\label{ratios}

Here, we present a discussion on the variations in the relative line strength and the line flux ratios, which can provide information on the gas conditions at different positions along the HH~1165 jet. Figure~\ref{goodman} shows the SOAR/Goodman spectra obtained at the source position and the knots marked in Fig.~\ref{SII-img}. The intensity in emission is expected to decline as the jet propagates to greater distances from the driving source. Thus the farthest knots in Fig.~\ref{goodman} show the weakest emission in all of the detected lines. Other than the [OI], [SII], and [NII] FELs, there is detection in the H$\alpha$ and H$\beta$ lines in both the northern and southern knots closer to the driving source. The H$\alpha$ and H$\beta$ lines are known to be strong tracers of extended outflow in HH objects (e.g., Reipurth \& Bally 2001). The jet knots are radiative shocks either internal to the jet material, or driven into the ambient medium. The H$\alpha$ and H$\beta$ emission arises from recombination in the cooling zone of the shock, and such shocks normally have strong Balmer line and continuum emission.

\begin{figure}[h]
\center
\includegraphics[scale=0.65]{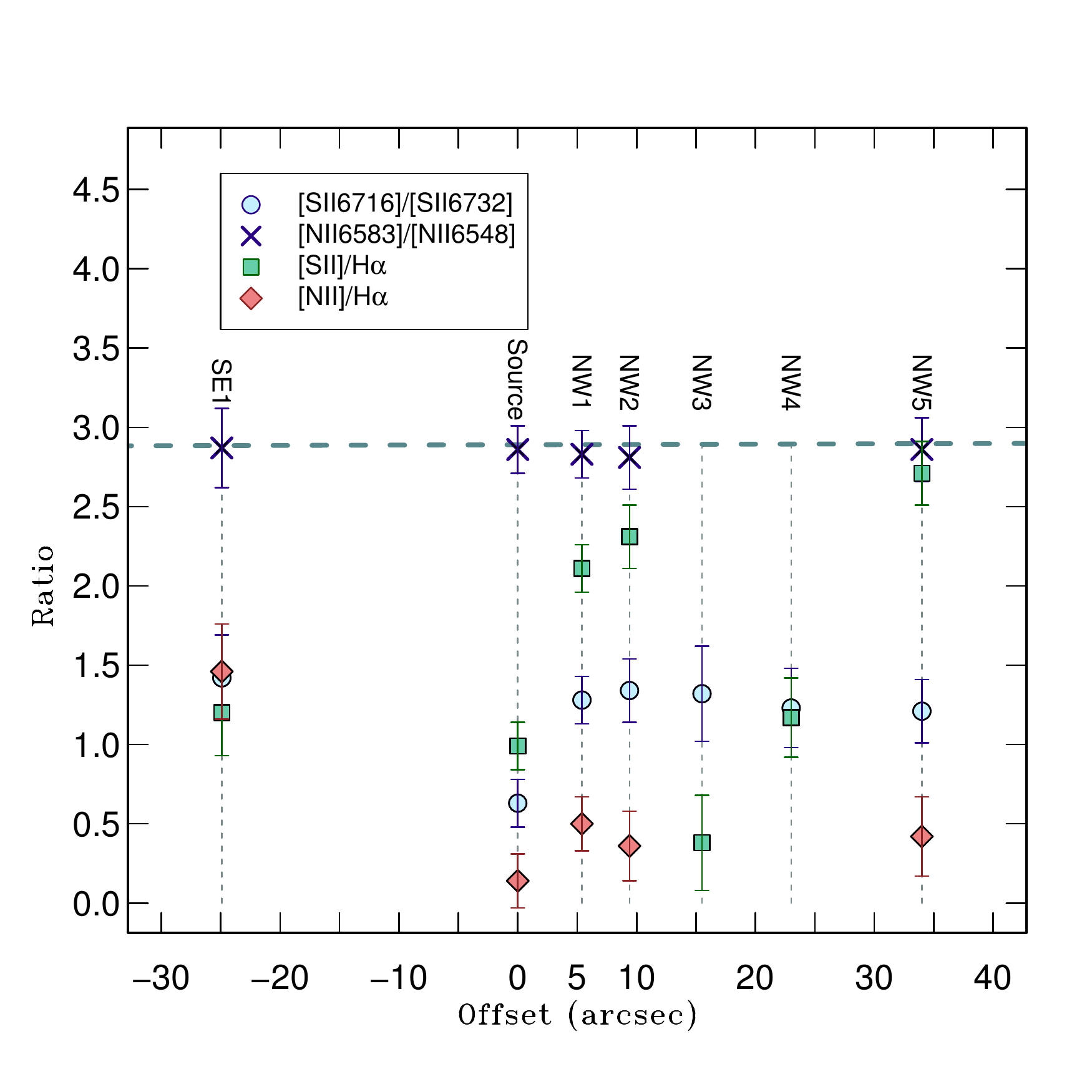}
\caption{Spatial variation of forbidden line ratios along the HH1165 jet and the M1701117 driving source. The horizontal dashed line corresponds to the theoretical [NII] 6583/6548 ratio. }
\label{line-ratios}
\end{figure}

Figure~\ref{line-ratios} plots the different line ratios for each knot and feature along the direction of the jet, using the data from the Goodman spectra. The line ratios are listed in Table~\ref{flux-ratios}. The individual spatial profiles of the [SII]$\lambda\lambda$6716,6731 and H$\alpha$ lines are shown in Figs.~\ref{strength-SII};~\ref{strength-Ha} (Sect.~\ref{appendix}. The [NII] (6583/6548) ratio is included in Fig.~\ref{line-ratios} and Table~\ref{flux-ratios} as a check on the quality of the line ratio estimates; our measured values agree well with the theoretical value of 2.93, shown by the horizontal dashed line in Fig.~\ref{line-ratios}, and do not suggest any systematic trend with position.

The optical spectra of low velocity shocks advancing into predominantly neutral gas are dominated by the forbidden lines of low ionization species, most notably [OI] and [SII], which are relatively enhanced compared to the Balmer lines and forbidden lines from higher ionization species such as [NII] (e.g., Hartigan et al. 1994). As the shock velocity increases, these low excitation forbidden lines become weaker relative to H$\alpha$. For shock velocities $>$ 90 km s$^{-1}$, UV photons emitted in the post shock cooling begin to ionize the gas ahead of the shock, due to which lines of higher ionization states such as [OII], [OIII] and [NII] become relatively stronger compared to the Balmer lines and the [OI] and [SII] forbidden lines. Conversely, the spectrum of a low velocity shock advancing into a medium which is already fully or partially ionized, e.g. by UV photons from early type stars, will mimic that of a higher velocity shock with relatively weaker [OI] and [SII] lines and stronger [NII] compared to the Balmer lines. Thus the [SII]/H$\alpha$ ratio can be used as a measure of shock velocity, while the [NII]/H$\alpha$ ratio provides information about the ionization state of the pre-shock gas. \footnote{We note that while [OI]$\lambda \lambda$6300,6363 line is also a good shock diagnostic, we were unable to unravel it from the strong terrestrial air glow emission in our low resolution spectra.} Correspondingly, while ``classical'' HH jets found in regions of predominantly low mass star formation, such as HH~34 and HH~111 (Reipurth et al. 2002; 1997), are bright in the [SII] and [OI] lines, externally ionized jets such as HH~502-506  (Bally \& Reipurth 2001) with comparable shock velocities have spectra with weak [SII] and [OI] that are much less distinct from that of the HII region into which they propagate.

The knots NW1, NW2, and NW5 in the HH~1165 jet have [SII]/H$\alpha$ $>$2 and [NII]/H$\alpha$ $<$0.5 (Fig.~\ref{line-ratios}; Table~\ref{flux-ratios}). The knot NW5 is the most extreme in this respect with a larger [SII]/H$\alpha$ line ratio than any other knot. Shock models covering a wide range of shock velocities, pre-shock densities, and magnetic field strengths predict that low shock velocities of $\sim$20-30 km s$^{-1}$ and very low pre-shock ionization fraction of $<$ 0.1 are required to explain such high values for the [SII]/H$\alpha$ ratio and the very small values for the [NII]/H$\alpha$ ratio (e.g., Hartigan et al. 1994; Nishikawa et al. 2007). This suggests that these knots represent low velocity shocks advancing into gas which is largely neutral and mainly shock-excited, and must be shielded from the ionizing radiation from the young stars in the region. The driving source has [SII]/H$\alpha$ $\sim$1 and [NII]/H$\alpha$ $\sim$0.1, which also suggests very low ionization effect in the ambient medium and a shock velocity of $\sim$40 km s$^{-1}$. In comparison, the SE1 knot shows a higher [NII]/H$\alpha$ ratio than any of the NW knots, indicating that the ambient gas in the southern part of the jet is considerably ionized. The [SII]/H$\alpha$ ratio for SE1 is $\sim$1, which suggests a shock velocity of $\sim$40 km s$^{-1}$ similar to what is predicted close to the driving source. Overall, we find a tentative trend where the [NII]/H$\alpha$ ratio decreases from the southern to the northern knots, suggesting that the ionization fraction of the ambient gas decreases north of the driving source. %This presents a puzzling picture wherein the NW part of the outflow appears to be a classical HH jet running into a predominantly neutral medium, while the SE beam resembles an externally photo-ionized jet. 

The episodic mass loss produces density variations along the length of the jet. We have calculated the electron densities, n$_{e}$, using the [SII] (6716/6731) flux ratio, and assuming an electron temperature of 8200 K (Hartigan et al. 1995). We have used the {\it stsdas/nebular} task in IRAF to calculate the electron densities. The bright emission in the [SII] doublet towards the source or the base of the jet implies an electron density of 3124 cm$^{-3}$, which decreases by more than a factor of 10 at the NW and SE knots (Table~\ref{flux-ratios}). The NW5 knot is denser than any of the other knots, with n$_{e}$ of 216 cm$^{-3}$, while the least dense is the SE1 knot (n$_{e}$$\sim$20 cm$^{-3}$). Thus the pre-shock density of the ambient gas is significantly low for SE1, but increases towards the northern knots. 

%The SE knots appear to be relatively closer to the massive star HR~1950, due to which stronger irradiation effects can be expected in the southern part of HH~1165 (Sect.~\ref{photoevaporated}). 

The shock velocities deduced from the [SII]/H$\alpha$ ratio are relatively low ($\sim$20-40 km s$^{-1}$) for most knots. An anomalous knot is NW3, which has the lowest [SII]/H$\alpha$ ratio of 0.38, suggesting a high velocity ($>$60 km s$^{-1}$) shock, based on the Hartigan et al. (1994) models. A closer look at the NW knots in Fig.~\ref{line-ratios} shows a zigzag pattern in the [SII]/H$\alpha$ ratio; it increases from $\sim$1 at the source to 2.3 at NW2, then drops to 0.38 at NW3, and then increases again to 2.7 at NW5. This suggests two possible episodic events in the jet. As noted in Sect.~\ref{SII}, NW3 has the shape of a partial bow shock and appears to be the head of a new outflow, while the farther NW4-8 could be part of an older outburst. The bending in the HH~1165 jet is also seen around NW3-4 knots. The zigzag pattern may be reflecting different epochs of outflow bursts (Sect.~\ref{deflected}).

%The rough estimates based on the shock models by Hartigan et al. (1994) are $\sim$20-30 km s$^{-1}$ for NW1, NW2, and NW5, and $\sim$40 km s$^{-1}$ for NW4 the driving source, and SE1. 

%At present, we do not have a good quality [NII] line detection for this knot. 

HH~1165 shows an asymmetric morphology, where the northern part is less ionized, brighter, and denser than the southern part of the jet. The possible causes for the jet asymmetry are discussed in Sect.~\ref{discussion}. We note that this was a preliminary spectroscopic study using low-resolution spectra obtained along a single slit orientation that only included the NW 1-5 and SE1 knots. The NW 6-8 knots are off the slit and the [NII] line detections are of a poor quality for the NW3 and NW4 knots and not plotted in Fig.~\ref{line-ratios}. The trends seen in Fig.~\ref{line-ratios} are thus tentative rather than definitive, and require higher resolution observations of each knot along the full length of the jet.

We have also measured the Balmer decrement or the H$\alpha$/H$\beta$ ratio for the source using the UVES spectra, and for the knots along the jet using the Goodman spectra (Table~\ref{flux-ratios}). The smallest value for the Balmer decrement is for the source (1.86 from UVES; 2.2 from Goodman). It ranges between 3 and 7 for the NW knots. The highest values for the Balmer decrement are found for the two inner most knots NW1 (7.7) and SE1 (6.7). The ratio H$\alpha$/H$\beta$ $\sim$2.0 for M1701117 is smaller than the intrinsic ratio of $\sim$2.8. We already know from the UVES spectral analysis for the driving source that there is a significant outflow component seen in the Balmer lines (Fig.~\ref{accretion_lines}). This could mimic the small Balmer decrements that are seen in very dense environments (for example, stellar winds), where the nebular approximation may not hold, i.e. decays occur much faster than excitations (e.g., Hartigan et al. 1994; 1990). The much steeper decrements for the knots compared to the source indicate reddening from dust. The Balmer decrement rises to $>$5 for the two inner most knots because those are in the regions most affected by the bright scattered light emission (reflection nebulosity and halo) seen in the H$\alpha$ image (Sect.~\ref{halpha-img}). The variation of the Balmer decrement for the knots is likely caused by differing decrements in the scattered light from the star HR~1950, and perhaps the intrinsic emission from the jet. HH1165 seems to be a case similar to HH30. The observed Balmer decrement in the bright knots in HH30 varies between $\sim$3.0 and 3.5, and increases to $>$5 in the reflection nebula (Hartigan \& Morse 2007).

%, and the electron temperature is expected to be $\sim$5000 K, lower than for nebular conditions from 10,000-20,000 K

%We note that the ratio of H$\alpha$/H$\beta$ is 2.1 for the driving source suggests a shock velocity of 80-100 km s$^{-1}$ (e.g., Hartigan et al. 1994). High shock velocities can be expected at the base of the jet close to the driving source, and it is likely that the H$\alpha$ emission from the jet is contaminated by the H$\alpha$ from the source. 

\section{Discussion}
\label{discussion}

\subsection{Asymmetries}
\label{photoevaporated}

HH~1165 is a peculiar case not because it shows an asymmetric morphology, which is more commonly seen in HH jets than a perfect symmetry, but because the characteristics of the northern and southern parts are very different. We find a puzzling picture wherein the NW part of the outflow appears to be a classical HH jet running into a predominantly neutral medium, while the SE beam resembles an externally photo-ionized jet. We have looked into the environs of the jet to explain the asymmetry. Figure~\ref{WISE} shows the wider surroundings of HH~1165 in the WISE mid-infrared image. An ionization front is seen very brightly on the eastern side of M1701117. HH 1165 is inside the IC434 HII region, which is centered on the $\sigma$ Orionis cluster core ($\sigma$ Ori AB). The radius of the IC434 HII region, i.e. the dense region where the bulk of the H$\alpha$ emission is concentrated, is $\sim$5 pc (e.g., Ochsendorf \& Tielens 2015; assuming a distance of 387 pc to the cluster). The distance between M1701117 and $\sigma$ Ori AB is $\sim$3.2 pc. The space motion of $\sigma$ Ori AB is directed eastward towards the L1630 molecular cloud, which is associated with the Horsehead nebula. The ionized gas in the IC434 HII region is due to the photo-evaporation of the L1630 cloud by the ionizing flux from $\sigma$ Ori AB. However, the ionizing flux from the $\sigma$ Ori core is not uniform throughout the HII region. If we consider the ionization parameter as calculated in Ochsendorf \& Tielens (2015), then the extent of ionization decreases with increasing distance into the HII region and moving away from the $\sigma$ Ori core. The ionization parameter is about 2 orders of magnitude lower at a distance of $\sim$3-3.5 pc compared to $\sim$1 pc from $\sigma$ Ori AB. The extent of ionization begins to increase again towards the L1630 cloud beyond $\sim$5 pc from $\sigma$ Ori AB. Thus, even though there are no neutral regions in IC434 as also noted from the mid-infrared spectra presented in Ochendorf \& Tielens (2015) and the gas is ionized, the extent of ionization is comparatively the lowest at a distance of $\sim$3-3.5 pc from the $\sigma$ Ori cluster core.

%such that the north-west part exhibits a classic HH jet, while the south-east part resembles an externally irradiated jet. 

\begin{figure}[t]
\centering
\includegraphics[scale=0.37]{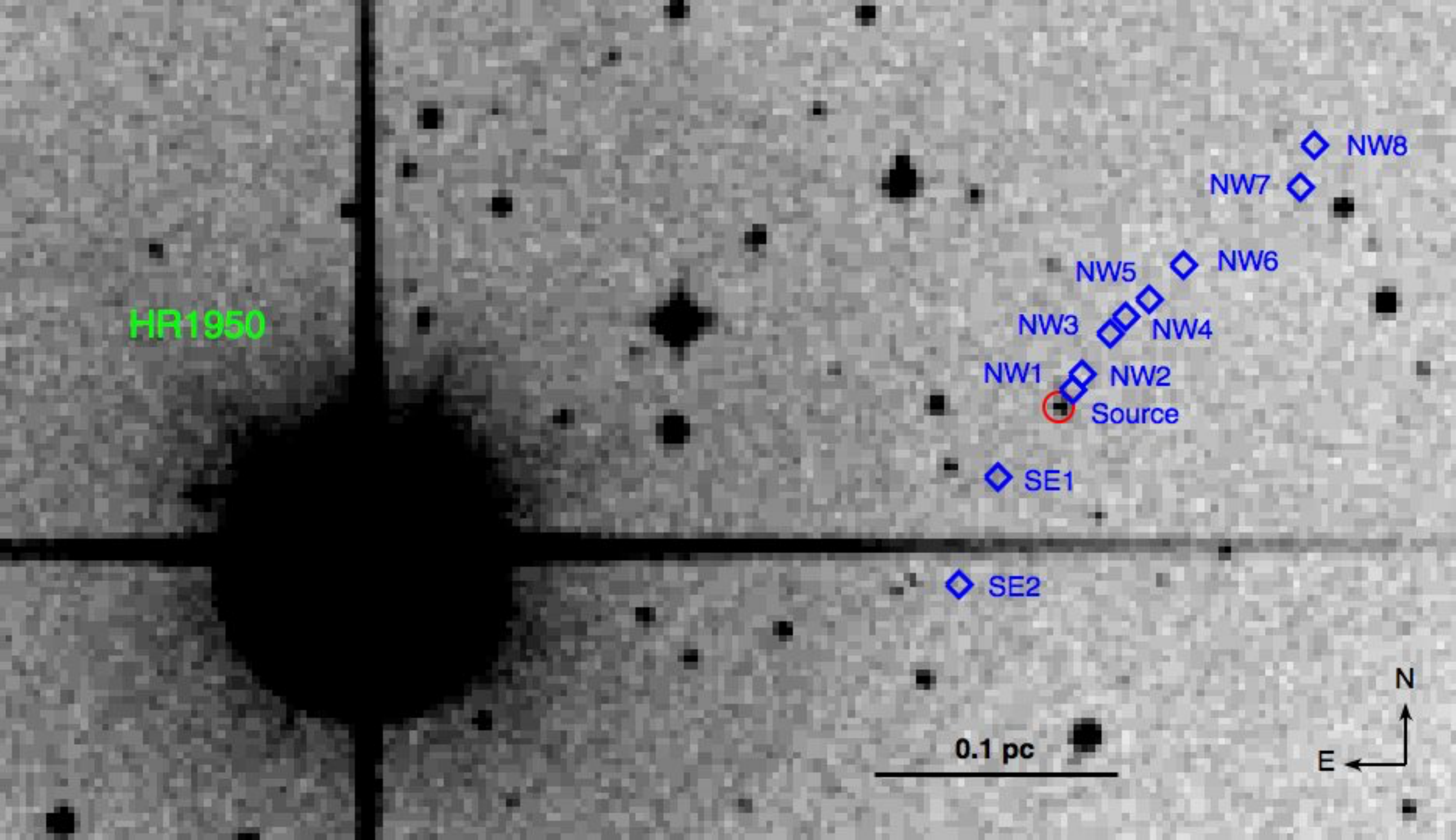}
\caption{The location of HH 1165 jet with respect to the B2 star HR~1950. The various knots seen in the jet are marked by blue diamonds. North is up, east is to the left. The image size is $\sim$3.5$\arcmin \times$5.3$\arcmin$.}
\label{position-2}
\end{figure}

Figure~\ref{position-2} shows the immediate environs of the HH~1165 jet. The driving source for HH~1165 lies in close proximity ($\sim$0.33 pc) to the massive star HR~1950. The southern part of the jet (SE1 and SE2) being tilted more towards HR~1950 may have been photo-evaporated to a large extent due to external irradiation, while the northern part being oriented away from HR~1950 has allowed the jet to propagate into a predominantly neutral medium.  As also noted in Sect.~\ref{ratios}, the large [NII]/H$\alpha$ and small [SII]/H$\alpha$ ratio for the SE1 knot indicate that the ambient gas in the SE region is mostly ionized, and support the criteria typically used for externally irradiated jets (e.g., Bally \& Reipurth 2003; Reipurth et al. 2010; Comer\'{o}n et al. 2013).

This may explain why we do not see strong effects of ionization in terms of a high [NII]/H$\alpha$ ratio in any of the northern knots, and perhaps there was a neutral medium surrounding the outflow. Or, there may have been local recombination at each NW knot such that the post-shock regions in the flow have been compressed enough to recombine, making them relatively neutral even if the surrounding medium is partly ionized. In comparison, the close proximity of the southern knots to HR~1950 may have caused photo-ionization of a neutral outflow, resulting in the high ionization seen for the SE knots in the HH~1165 jet (Fig.~\ref{position-2}). However, the projected distance from HR~1950 to the southern most SE2 knot is $\sim$0.27 pc, and is similar to the $\sim$0.34 pc distance to the inner NW1-5 knots. Therefore, the effects of the external irradiation from HR~1950 and the extent of photo-ionization are expected to be similar on the northern and southern parts of the jet. The relative neutrality of the northern part of the jet is indeed a puzzle, and we aim to address it in the future with high resolution observations and investigate the presence of a local ionization front in each knot.

It is interesting to note that all of the previously known HH jets in $\sigma$ Orionis, HH~444-447 (Reipurth et al. 1998) and HH~1158 (Riaz \& Whelan 2015), are externally irradiated jets that show an asymmetric morphology, with one weak lobe due to irradiation effects from the $\sigma$ Orionis cluster core (Fig.~\ref{position}). Note that the most distant southern knot SE2 in the HH~1165 jet is located $\sim$51$\arcsec$ from M1701117, while the farthest northern knot NW8 is $\sim$91$\arcsec$ away (Fig.~\ref{SII-img}). It may be the case that both jet components were nearly equidistant from the driving source and HH~1165 had a symmetric morphology, but we have lost this evidence due to the effects of external radiation on the southern lobe.

\begin{figure}[h]
\center
\includegraphics[scale=0.6]{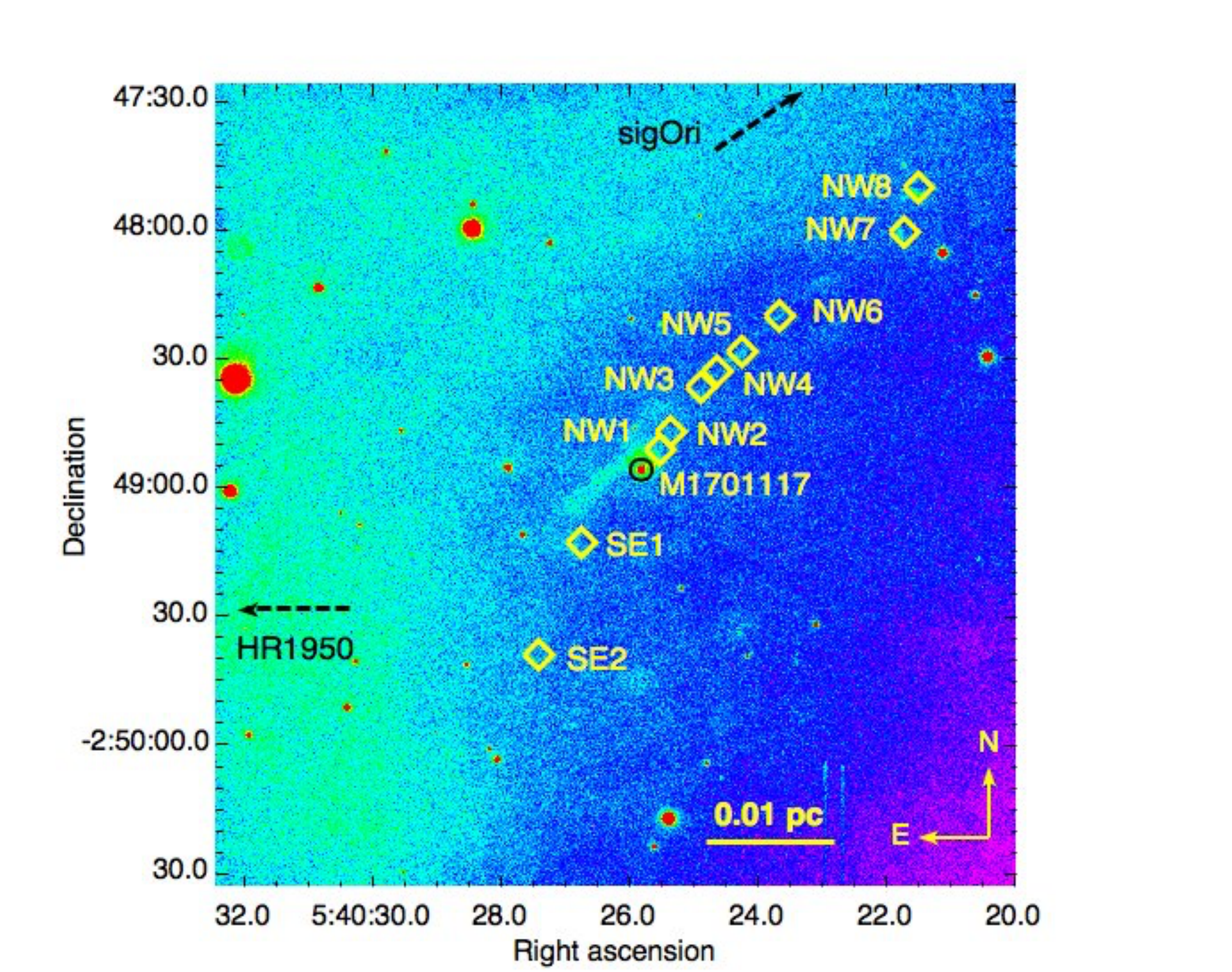}
\caption{The full 3$\arcmin \times$3$\arcmin$ H$\alpha$ image, showing the surroundings of the HH~1165 jet. Diamonds mark the various knots seen in the [SII] image, and the circle marks the driving source. The arrows point towards $\sigma$ Ori AB and HR~1950.  }
\label{Ha-full}
\end{figure}

The HH~1165 jet is highly asymmetric in brightness, with even the faintest north-west component being at least $\sim$4 times brighter (and denser) than SE1 or SE2 (Table~\ref{flux-ratios}). The asymmetry in brightness could be due to the density gradients in the surroundings of the HH~1165 jet. As is evident in Fig.~\ref{Ha-full}, there is a clear density (or intensity) gradient between the north-east and south-west sides of the source. As a rough estimate, the flux density in a 0.5$\arcmin$ square area in the north-east corner of the 3$\arcmin \times$3$\arcmin$ H$\alpha$ image is a factor of $\sim$1.8 higher than in the south-west corner. The one-sided cavity is likely caused by the density gradient (Fig.~\ref{Ha-img}). The eastern side appears denser and so the eastern cavity is also brightly seen, while the lack of material to the west of the jet could explain why we do not see much reflection nebulosity tracing the western cavity. A similar reasoning could be applied to the weak detection of the red-shifted lobe that may still be embedded in the natal material. In Fig.~\ref{Ha-full}, we also see a difference in the density in the north-west and south-east sides of the source, such that the knots SE1 and SE2 are located in denser regions than any of the northern knots. Hence, it may be the case that the south-east counter-jet is more embedded and still shrouded from us in the optical images.

Inclination effects are also important. A near pole-on inclination would imply that we are looking down the ``throat'' of the outflow, and thus only one cavity would be visible. The classic bow-tie shaped bipolar cavities observed in a few rare cases, such as, the low-mass protostar IRAS 04302+2247 (e.g., Padgett et al. 1999), indicate that the YSO must have a fairly high line-of-sight inclination angle (nearly edge-on disk). Based on the radiative transfer modeling of the optical to sub-millimeter spectral energy distribution (SED) for M1701117, the best model fit indicates an intermediate inclination of $\sim$50-60$\degr$ for the system (Riaz et al. 2015). For intermediate inclinations, a partial obstruction of the material directed away from the observer would be expected. This is consistent with the fact that we only see the blue-shifted lobe of the jet that begins to appear as close as $\sim$1.5$\arcsec$ ($\sim$0.003 pc) from the driving source. In contrast, the red-shifted lobe begins to appear at distances $>$10$\arcsec$ ($>$0.02 pc) from the source. The estimated size of the envelope+disk components for M1701117 from SED modeling is $\sim$1700 AU ($\sim$0.01 pc; Riaz et al. 2015). The partial obstruction is consistent with the intermediate inclination. The inner $\sim$0.02 pc of the red-shifted lobe is hidden by the envelope/disk, while the observed blue-shifted lobe could be pointing out of the cloud. Likewise, when only one cavity is observed, the back side cavity has likely been obscured beyond detection by high extinction from the circumstellar envelope/disk material.

\subsection{Deflected jet, wide outflow cavities, and multiple knots}
\label{deflected}

The bending seen in the HH 1165 jet is the characteristic {\bf C}-shaped symmetry observed previously in several HH outflows (e.g., Bally \& Reipurth 2001). One of the most spectacular examples of {\bf C}-shaped jets is the recently discovered HH 1064 driven by the young star Parenago 2042 in NGC 1977 in Orion (Bally et al. 2012). There are two main kinds of {\bf C}-shaped bent jets: one that show outward bending away from the cluster core, indicating deflection by a side wind, while the second show inward bending towards the cluster center, indicating that these stars may have been dynamically ejected from the cluster. Interestingly, a large number of {\bf C}-shaped HH jets in the Orion Nebula, most notably HH 502, show outward bending, and lie at large projected distances from the core of the cluster (e.g., Bally \& Reipurth 2001; Bally et al. 2001; 2006). In contrast, the inward bent jets, most of which have been observed in NGC 1333, lie well within the projected boundary of the cluster (e.g., Bally \& Reipurth 2001).

There are at least four mechanisms that can produce {\bf C}-shaped flows: (1) Motion of the source through the medium; (2) A side-wind; (3) Photo-ionization of a neutral outflow and the resulting Oort-Spitzer rocket effect; and (4) Radiation pressure from a nearby star.

For the case of HH 1165, the driving source lies at the periphery of the $\sigma$ Orionis cluster at a $\sim$3.2 pc distance from the cluster core (Fig.~\ref{position}). The symmetry of this jet appears to be similar to the outward bent cases, deflected away from the $\sigma$ Orionis core due to an outward-oriented force, such as a side wind. Masciadri \& Raga (2001) have shown from modeling of the bent HH 505 jet in Orion Nebula that the interaction of an externally ionized jet with a side wind can produce a cocoon that surrounds the jet and its driving source. This can be a cocoon inflated by a stellar wind or a cavity produced by the sideways splashing of a time-variable jet. We do see signs of a cavity surrounding the HH 1165 jet and its driving source, particularly towards the western side of the jet (Fig.~\ref{Ha-full}). The curvature of the jet neatly follows the apparent outline of this evacuated region, where we begin to see a change in the intensity. This may be the effect of the wide-angle winds from the $\sigma$ Orionis cluster core that have cleared a region around the jet and its surroundings. The presence of a wide-angle side wind acts like an outward oriented force, and a jet flowing into this wind will be decelerated and bent away from the core. Thus, mechanism (2) is possible.

However, the brighter blue-shifted jet beam in HH 1165 is oriented towards the $\sigma$ Orionis core (Fig.~\ref{position}), unlike the other outward bent jets in ONC that show jet beams oriented away from the cluster core. Another explanation for the jet deflection could be that the driving source M1701117 is located quite close ($\sim$0.3 pc) to the massive star HR~1950, and the red lobe in the HH 1165 jet is oriented towards this star (Fig.~\ref{position}). Stellar winds from HR~1950 collide with the outflow and produce the observed jet deflection. There is also a slight offset seen in the red lobe closer to the driving source (Fig.~\ref{comb-img}), which could also be due to a deflection caused by the wind-outflow collision. Even though early B-type stars are known to have weak winds, with the wind terminal speeds not significantly exceeding the stellar escape speed (e.g., Oskinova et al. 2011), the close proximity to HR~1950 must have an effect on deflecting the southern lobe of the jet, if not the most distant NW knots. Thus, mechanism (3) could be applicable here.

The bending in this jet could instead be due to the ionization front seen very brightly towards the NE of the jet in the WISE mid-infrared image (Fig.~\ref{WISE}). Also marked in the image are the locations of some known B-type stars [HD37806 (B9); HD37927 (B8III); HD37886 (B8III); HD294304 (B5)] close to M1701117 that may be producing the ionization front. There may also be a tussle between these multiple effects. The core of the $\sigma$ Orionis cluster being composed of several OB type stars will be more dominant in dissipating the material towards the western side and creating a cocoon around the jet, the ionization front towards the eastern side could push material forward towards the outflow/jet system, and produce a parabolic front or a sharp edge of dense material, as seen in the H$\alpha$ image (Fig.~\ref{Ha-full}), where the wind interacts with the debris that surrounds the jet (e.g., Bally et al. 2000; Bally \& Reipurth 2001). The deflection in such a scenario is caused by the interaction of the ambient medium directly with the jet, and the momentum is transferred either directly or indirectly from the stellar winds through the jet cocoon to the body of the jet (e.g., Masciadri \& Raga 2001). Thus, mechanism (4) could be the main cause of the jet deflection.

%To explore this further, Fig.~\ref{WISE} shows a wider 1$\times$1 degree view of the surroundings of the HH~1165 jet in the WISE mid-infrared 3.4$\mu$m image. 

The latest measurement on the proper motion for M1701117 is $\mu_\alpha$ cos $\delta$, $\mu_\delta$ = -2$\pm$2, -6.3$\pm$1.9 mas/a (Sect.~\ref{source}). As shown by Ochendorf \& Tielens (2015), $\sigma$ Ori AB is moving north-east at PA$\sim$60$\degr$ towards the L1630 cloud. The large projected distance of M1701117 from $\sigma$ Ori and its proper motion imply that it is likely moving towards the south-east with respect to $\sigma$ Ori (Fig.~\ref{WISE}), assuming that it is a member of $\sigma$ Orionis. Such a motion would also explain the {\bf C}-shaped bending in the jet, and mechanisms (1) and (2) could be acting in the same direction. Since M1701117 is in the interior of the IC434 HII region (Sect.~\ref{photoevaporated}), the flow of the plasma off the western wall of the Orion B cloud may produce a side-wind. As noted earlier, the deflection could also be caused by the side-wind from $\sigma$ Ori AB. The {\bf C}-shaped symmetry in HH 1165 could be produced by the combined effects from the massive stars in the ionization front to the east and the $\sigma$ Orionis cluster core to the west (Fig.~\ref{WISE}). However, the close proximity to HR~1950 (Fig.~\ref{position-2}) would imply that this B-type star dominates the Lyman continuum and total radiation fields, in which case mechanisms (3) and (4) will reinforce mechanism (2). If $\sigma$ Ori dominates, then mechanism (2) will be opposed by (3) and (4).

It is interesting to note that in the scenarios where (1) and (2) mechanisms are possible, and the ionization front towards the north-east side could push material forward towards the jet, the ``cavity walls'' traced by H$\alpha$ would be predominantly seen along the north-east because this is where the medium is likely to be most compressed. The south-west, trailing side would have lower density. In this sense, the morphology resembles the LL-Ori class of objects, which are deflected flow cocoons (e.g., Bally et al. 2000; Bally \& Reipurth 2001).

As mentioned, there is a slight offset seen in the southern part of the HH~1165 jet closer to the driving source (Fig.~\ref{comb-img}). This hints towards a possible mis-alignment between the disk rotation axis and the outflow axis, which could produce an off-axis shock. Such mis-alignments have been seen in Class 0/I sources where the magnetic field in the protostellar core is not aligned with the outflow (e.g., Hull et al. 2013), and in cases where the central driving source is a binary and the circumbinary disk is mis-aligned with the outflow (e.g., IRS43; Brinch et al. 2016). M1701117 could be an unresolved binary where either the disks around the individual components or the circumbinary disk is mis-aligned with the outflow. The companion to M1701117 may have ejected the southern lobe at a different velocity and at a higher angle than the primary flow. %It is an interesting scenario that can be probed further with high-resolution imaging. 

With regards to the extended outflow cavities observed in HH 1165 (Fig.~\ref{Ha-img}), where the south-east cavity extends to as far as $\sim$0.04 pc from the source, it has been suggested, based on X-wind model (Shu et al. 2000), that the presence of both a collimated jet and a broader wide-angle protostellar wind could carve out such wider outflow cavities over time (e.g., Konigl \& Pudritz 2000; Delamarter et al. 2000; Matt et al. 2003). According to this model, during the very early Class 0 stages, the dense, high-velocity, collimated component of the outflow can pierce through the high-density circumstellar envelope and evacuate the polar regions, allowing the spherical outflow to expand. Over time, the envelope loses mass, the infall rate decreases, and the outflow begins to dominate over the infall. This allows the weaker, less dense, and lower velocity broad-angle wind component to carve out a wider outflow cavity as the system evolves. Presently, the M1701117 system is in the Class I stage. The dynamical timescale of the jet, i.e. the time it must have taken for the jet to reach its apparent NW end at about $\sim$0.17 pc, assuming a tangential velocity of $\sim$100 km s$^{-1}$, is $\sim$1660 yr. On the other hand, the knot NW3 has a partial bow shock structure facing North, and it has a nice trailing wing in H$\alpha$ (Fig.~\ref{SII-img}). Perhaps this could be the head of the outflow from a new outburst traveling along the NW1-3 axis, with the knots further out an older flow with a slightly different direction. Thus the dynamical age of this jet may be younger than a thousand years. In comparison, small-scale micro-jets with spatial extents of a few arcseconds have dynamical timescales of a few hundred years (e.g., McGroarty \& Ray 2004). This system is in a relatively more evolved stage where the low-velocity wind has by now carved wider and spatially extended outflow cavities. %The system is also young enough with some dusty material left in the cavities to produce the observed bright nebulosity.

The morphology of large-scale outflows can provide a fossil record of the mass-loss history of their parent stars, considering their tens of thousands of years long dynamical timescales. The presence of several knots or segments of ejecta separated by relatively empty gaps indicates episodic events or a variable outflow for HH 1165. We do not, however, see expanding knots farther from the source and this could be due to high cooling efficiency, such that the jet hits the cloud material but cools down quickly before it expands. This is unlike most wide-length HH jets in Orion for which many of the knots, particularly the more distant ones, appear to form mini bow shocks that extend into the surrounding medium (e.g., Reipurth \& Bally 2001). For HH 1165, it is only the NW3 knot that shows a clear bow-shock morphology, while the rest appear elongated along the direction of propagation of the jet (Fig.~\ref{SII-img}). It appears that the NW4-NW7 and SE1-2 knots are amongst the oldest ejecta, while NW1-3 being the closest to the driving source, is the latest to be ejected. The knots in HH 1165 are also not equidistant; while NW3-NW5 appear to have been ejected at regular intervals, there is a $\sim$10$\arcsec$ separation between NW2 and NW3. For large-scale outflows that have had time to evolve, the extended shocks have undergone multiple interactions with the ambient medium and the previous, slower ejecta. Many of these shocks, particularly the more distant ones, fade quickly with time, and only the strongest and the largest shocks remain. Due to this, the gap between the consecutive knots increases with distance from the driving source. The overall morphology of the knots thus appears irregular and complex.

Precession of the outflow may also explain the directional change in the jet (e.g., Raga et al. 2001; Masciadri \& Raga 2002). The models by e.g., Raga et al. (2001) have shown that a precession of the outflow axis can produce off-center positions for the jet beam, but also result in a broadening of the successive working surfaces as the flow moves away from the driving source, such that the width of the farthest knot is comparable to the opening angle or the local width of the precession cone. As noted in Sect.~\ref{SII}, HH 1165 shows a very collimated jet, with the width of the farthest knot similar to the one closest to the driving source. It is also argued that if precession is causing the deflection then it should result in a repeated pattern of the observed shape, rather than a single large area deflection. The jet axis precession has also been suggested to produce wide cavities over time. However, in most cases, the angle of the precession is too small to account for the full extent of the outflow cavity widening (e.g., Reipurth et al. 2000). Therefore, the case of a precessing jet does not seem applicable to the features observed for HH 1165.

\subsection{Comparison with large-scale HH jets}

The uniqueness of the HH 1165 jet lies in the fact that it shows all of the well-known morphologies, such as, the {\bf C}-shaped bent jet, multiple shock emission knots, a bow shock, and bright reflection nebulosity, which have been seen among high- and low-mass protostars driving powerful HH jets in Orion (e.g., HH 502/503, HH 201, HH 212, HH 31, HH 34). The emission line spectra seen along the length of the HH~1165 jet is also typical for such jets, with [SII] bright knots characteristic of low velocity internal shocks along the jet beam and regions with stronger H$\alpha$ in the bow shocks. Over the last two decades, the advent of large CCD mosaics has resulted in the discovery of several large-scale HH flows stretching lengths of several parsecs. Some notable examples among giant flows with projected lengths greater than 1 pc and reaching beyond 10 pc are HH 34, HH 131, HH 1, HH 2, HH 401/402, HH 46/47, HH 113/HH 111/HH 311 (e.g., Reipurth et al. 1997; Bally \& Devine 1994; Wang et al. 2005; Ogura 1995). A majority of these large HH flows are driven by low-mass ($\sim$0.3--1 $M_{\sun}$) embedded Class 0/I sources. Parsec-scale outflows have also been detected in a few Class II low-mass stars, which were previously only associated with small-scale micro-jets of $\leq$0.03 pc length (e.g., McGroarty \& Raga 2004). Among very low-mass stars and brown dwarfs ($\sim$0.08--0.2 $M_{\sun}$), there are now a number of Class II sources that are known to drive jets (e.g., Whelan et al. 2014). However, all of these are cases of micro-jets, similar to those driven by T Tauri stars, with projected lengths of $\sim$0.002-0.01 pc (Table~\ref{known-jets}).

An evolution in the outflow length with the evolutionary stage of the driving source has been noted among protostellar jets in the Orion A cloud (e.g., Stanke et al. 2002). During the very early Class 0 stage of evolution, the outflow expands quickly from an initial zero length to a maximum parsec scale extent, and about one-third of the jets have parsec scale lengths. Then, during the Class I stage, the jets begin to shrink again, and have typical lengths of about 1/2 parsec, with just about 10\% being parsec scale jets. The jets shrink further during the more evolved Class II stage, with mean lengths of only $\sim$0.2 pc. There can be anomalous cases, such as the HH jets in $\sigma$ Orionis driven by T Tauri stars, where photo-evaporation due to external radiation has shrunk the jet length to less than 0.01 pc (Reipurth et al. 1998). But the overall evolution in the jet length suggests that most if not all outflows go through a parsec scale phase. While the present observations on jets in very low-mass stars and brown dwarfs are still very sparse, we find a possible trend in the evolution of the jet length with the evolutionary stage of the driving source, from M1701117 (Class I) with a $\sim$0.26 pc length to Class Flat and Class II driving sources with jets of $\leq$0.01 pc length (Table~\ref{known-jets}).

We note that young stars evolve in close synchrony with their surrounding medium, as the ambient material from the parent cloud is cleared by the combined effects of mechanical momentum injected by winds and outflows, and photoionization from the massive O and B stars of the association. Therefore, not only do the ejecta become weaker with time, as stars evolve from the embedded Class 0/I phase to optically visible Class II T Tauri stars, but also the outflows and jets have much less material to interact with as most of it is being dissipated. The apparent decrease in outflow size during the Class II stage may be a selection effect or artifact of decreasing mass-loss rate, which implies that ejecta have lower densities at a given distance, making them much fainter that the ejecta from Class 0/I sources with higher mass-loss rates.

The discovery of HH 1165 shows that outflows can attain such large lengths even among very low-luminosity objects. This can be expected, since their tangential velocities are typically in the range of $\sim$50-100 km s$^{-1}$, and the outflow phase for very low-mass stars and brown dwarfs, both in the Class I and II phases, can last for at least 0.1-1 Myr (e.g., Whelan et al. 2014; Riaz \& Whelan 2015). Such jets may be rare due to a lack of the environmental conditions needed to allow the jet to propagate to, and still be visible at large distances from the source. The need for deep sensitive observations for detection of such faint jets may also be an obstacle to their discovery. We would expect that low luminosity sources are more likely to be found in low density environments, after disruption/fragmentation of very low-mass cloud cores. So the problem is more likely lack of dense material to shock against than the difficulty of propagation, which should be easier in lower density medium. Indeed, Figs.~\ref{WISE};~\ref{position} show that being the closest to the dense region towards the east, and farthest from the $\sigma$ Orionis cluster core towards the west, has provided dense enough material to produce shock emission, as well as allowed the HH~1165 jet to propagate to much larger lengths than any other HH jet in this region. The analogies in the morphology and the various features seen in the HH~1165 jet and protostellar jets are encouraging, and indicate that the jet launching models, such as, the disk wind and the X-wind models (Sect.~\ref{deflected}), proposed for jets in low- and intermediate-mass stars are also applicable for objects at the sub-stellar limit. The discovery of HH 1165 provides the first evidence that large HH flows can be driven by objects at all stellar masses, and appear to be a ubiquitous phenomenon. HH 1165 shows all of the signatures to be considered as a scaled-down version of parsec-length HH jets, and can be termed as the first sub-stellar analog of a protostellar HH jet system.

\acknowledgments

We thank the referee, John Bally, for his detailed review of the paper. We are grateful to Bob O'Dell and M. Tafalla for an in-depth discussion on the jet. We thank J. Caballero for providing an estimate on the proper motion of the source. BR acknowledges funding from the Marie Sklodowska-Curie Individual Fellowship (Grant Agreement No. 659383). Based on observations obtained at the Southern Astrophysical Research (SOAR) telescope, which is a joint project of the Minist\'{e}rio da Ci\^{e}ncia, Tecnologia, e Inova\c{c}\~{a}o (MCTI) da Rep\'{u}blica Federativa do Brasil, the U.S. National Optical Astronomy Observatory (NOAO), the University of North Carolina at Chapel Hill (UNC), and Michigan State University (MSU). Based on observations collected with UVES at the Very Large Telescope on Cerro Paranal (Chile), operated by the European Southern Observatory (ESO).

%\bibliographystyle{apj}
%\bibliography{M170-V2_ref}

\clearpage

\appendix
\label{appendix}

\counterwithin{table}{section}

\section{Tables}

\begin{deluxetable}{cclccc}
	\tabletypesize{\scriptsize}
	\tablewidth{0pt}
	\tablecaption{SOAR SAM Observing Log \label{tabobs}}
	\tablehead{
		\colhead{UT Date}    & \colhead{UT(begin)}& \colhead{Filter} & \colhead{Exposure} & \colhead{No. of exposures} & \colhead{Airmass range} \\
		\colhead{yyyy-mm-dd} & \colhead{hh:mm}    & \colhead{}       & \colhead{s}        & \colhead{}                 & \colhead{}
		}
	 \startdata
	 2015-12-15              &  05:34             &   R(Bessell)     &   300              &  5                         &   1.15-1.17  \\
	 2015-12-15              &  06:02             &   H$\alpha$      &  1200              &  3                         &  1.19-1.28   \\
	 2015-12-15              &  07:28             &     [SII]        &  1200              &  3                         &  1.45-1.72   \\
	 2015-12-17              &  03:44             &   H$\alpha$      &  1200              &  3                         &  1.17-1.13   \\
	 2015-12-17              &  04:49             &     [SII]        &  1200              &  3                         &  1.13-1.15   \\
	\enddata
\end{deluxetable}

\newpage

\begin{longtable}{rp{1.5cm}p{2.5cm}rp{0.5cm}rp{2.cm}} 
\caption{Fluxes and equivalent widths of lines identified in the UVES spectra}\\
\hline \hline 
Line & $\lambda_{central}$  & Line Flux  & Equivalent Width    \\ 
         & ({\AA})   & (10$^{-16}$ erg cm$^{-2}$ s$^{-1}$) &  ({\AA})  \\
\hline

H$\zeta$        			& 3889.05  & 3.9$\pm$0.3 & $-$43.3$\pm$30    \\
Ca~{\sc ii} K   			& 3933.66 & 8.2$\pm$1.0  & $-$18.2$\pm$1.0    \\
H$\epsilon$ + Ca~{\sc ii} H  	& 3970.07 & 9.1$\pm$0.5  & $-$34.7$\pm$0.5     \\ \relax
[S~{\sc ii}]  			& 4068.6 & 38.8$\pm$2.0  & $-$50.8$\pm$10.0     \\  \relax
[S~{\sc ii}]  			& 4076.3 & 12.6$\pm$0.3  & $-$21.2$\pm$1.0    \\
H$\delta$       			& 4101.74 & 6.5$\pm$0.6  & $-$30.0$\pm$3.0     \\ 
Fe~{\sc iii}				& 4243.75 & 2.7$\pm$0.5 & 10.5$\pm$1.0 \\  \relax
[Fe~{\sc ii}]			& 4287.23 & 2.8$\pm$0.5 & 9.1$\pm$1.0 \\
H$\gamma$      		& 4340.47 & 9.2$\pm$0.2  & $-$24.0$\pm$3.0     \\
Fe~{\sc i}				& 4814.32 & 1.5$\pm$0.1 & 1.9$\pm$0.2 \\
H$\beta$            		& 4861.33 &  18.8$\pm$0.5  & $-$21.7$\pm$2.0     \\  \relax
Na~{\sc i}				& 5890.05 & 5.4$\pm$0.3 & 10.4$\pm$1.0 \\
Na~{\sc i}				& 5896.04 & 5.9$\pm$0.5 & 11.8$\pm$1.0 \\   \relax
[O~{\sc i}]           		& 6300.30 &  29.7$\pm$0.4  & $-$39.3$\pm$2.0    \\    \relax
[O~{\sc i}]          		& 6363.78 & 10.0$\pm$0.4 & $-$13.9$\pm$1.5   \\   \relax
[N~{\sc ii}]    			& 6548.0 & 0.9$\pm$0.2  & $-$0.9$\pm$0.2    \\
H$\alpha$          		& 6562.85 &  39.5$\pm$1.5 & $-$40.4$\pm$4.0    \\  \relax
[N~{\sc ii}]                      	& 6583.45 &  4.5$\pm$0.5  & $-$5.4$\pm$1.0    \\   \relax
[S~{\sc ii}]                      	& 6716.44 &  20.2$\pm$1.0 & $-$26.2$\pm$2.0    \\   \relax
[S~{\sc ii}]                      	& 6730.82 &  36.3$\pm$0.6 & $-$38.0$\pm$2.0    \\ 
He~{\sc i}				& 7064.80 & 0.3$\pm$0.1 & 0.3$\pm$0.1 \\   \relax
[Fe~{\sc ii}]			& 7155.17 & 6.3$\pm$0.5 & 7.9$\pm$1.2 \\   \relax
[Fe~{\sc ii}]			& 7171.99 & 1.7$\pm$0.2 & 2.0$\pm$0.3 \\   \relax
[Ca~{\sc ii}]			& 7291.47 & 4.3$\pm$0.6 & 4.9$\pm$1.0 \\   \relax
[O~{\sc iii}]           		& 7319.65 &  4.4$\pm$0.3  & $-$4.9$\pm$0.5    \\    \relax
[O~{\sc iii}]           		& 7323.71 &  4.3$\pm$0.2  & $-$5.5$\pm$0.5    \\    \relax
[O~{\sc iii}]           		& 7329.85 &  3.4$\pm$0.4  & $-$4.0$\pm$0.8    \\    
Th~{\sc i}				& 7377.62 & 2.6$\pm$0.2 & 2.8$\pm$0.2 \\   \relax
[Fe~{\sc ii}]			& 7388.16 & 1.3$\pm$0.1 & 1.4$\pm$0.1 \\
Fe~{\sc i}				& 7411.15 & 0.6$\pm$0.4 & 0.6$\pm$0.3 \\
Th~{\sc i}				& 7452.4 & 2.0$\pm$0.2 & 2.5$\pm$0.3 \\
Fe~{\sc i}				& 8125.02 & 0.5$\pm$0.3 & 0.5$\pm$0.3 \\
Ca~{\sc ii}                    	& 8498.02  &  1.4$\pm$0.2 & $-$1.5$\pm$0.2    \\
Ca~{\sc ii}                    	& 8542.09  &   2.6$\pm$0.2 & $-$2.9$\pm$0.4   \\  \relax
Ne~{\sc ii}				& 8578.41 & 0.6$\pm$0.1 & 0.6$\pm$0.1 \\
Fe~{\sc i} 				& 8616.21   &  4.7$\pm$0.5 & $-$5.3$\pm$0.5    \\
Ca~{\sc ii}                    	& 8662.14  &  1.8$\pm$0.2 & $-$1.8$\pm$0.2    \\   \relax
[C~{\sc i}]  			& 8726.95  &  0.7$\pm$0.1 & $-$0.7$\pm$0.1    \\
Fe~{\sc i}				& 8891.67 & 2.2$\pm$0.2 & 2.4$\pm$0.1 \\   \relax
[Fe~{\sc ii}]			& 9033.09 & 0.8$\pm$0.3 & 0.8$\pm$0.3 \\
O~{\sc ii}				& 9051.42 & 1.2$\pm$0.2 & 1.3$\pm$0.2 \\
Fe~{\sc i}				& 9068.49 & 0.9$\pm$0.3 & 1.0$\pm$0.3 \\   \relax
H~{\sc i} (Pa 9) + [Fe~{\sc ii}]    	& 9229.15  &  1.3$\pm$0.2 & $-$1.5$\pm$0.2    \\   \relax
[Fe~{\sc ii}]			& 9267.05 & 0.9$\pm$0.3 & 1.1$\pm$0.2 \\
\hline\hline      
\label{fluxes}
\end{longtable}

\newpage

\begin{table}[h]
\center
\caption{Mass accretion rate measured using different diagnostics for M1701117}
\begin{tabular}{ccc}
\hline
\hline
$log$ ($\dot{M}_{acc}$)  &  delta  & Line  \\
\hline
	
-9.30E+00  &  2.70E-01  &  Ca II (K) \\
-8.95E+00  &  2.30E-01  &  Ca II (H) \\
-9.13E+00  &  2.80E-01  & H$\delta$ \\
-9.13E+00  &  2.50E-01  & H$\gamma$  \\
-9.15E+00  &  2.30E-01  &  H$\beta$  \\	
-9.91E+00  &  2.61E-01  &  H$\alpha$ \\	
-8.77E+00   &  3.20E-01 &  NaI  \\
-8.59E+00   &  3.70E-01 &  NaI  \\
-8.74E+00   &  4.10E-01  &  [OI]$\lambda$6300  \\
-9.49E+00   &  3.91E-01   &  HeI  \\
-1.00E+01   &  3.81E-01   & CaII (8498 {\AA}) \\
-9.91E+00   &  4.21E-01   &  CaII (8542 {\AA})  \\
-9.87E+00   &  4.32E-01   &  CaII (8662 {\AA})  \\
\hline             
\end{tabular}
\label{acc-rates}
\end{table}

\begin{table}[h]
\center
\caption{Line flux ratios measured for each feature along the jet}
\begin{tabular}{p{2cm}cp{2cm}cp{1.5cm}cp{2cm}cp{1.5cm}cp{1.5cm}cp{1.5cm}}
\hline
\hline
Feature  		& [SII]/H$\alpha$ & [NII]/H$\alpha$ 	& 	[SII] 		& n$_{e}$($\times$10$^{2}$)		& [NII]  		& H$\alpha$/H$\beta$ \\
              		&   			   &			      	&  (6716/6731)	& [cm$^{-3}$]	                                  & (6583/6548)   & \\
\hline
M1701117     	&	0.99  	   & 	0.14		      	&    	0.63 		& 30$\pm$10  		& 	2.86  & 2.2 	\\
NW1     		&    	2.11   	   &   	0.50  		&    	1.28    	& 1.4$\pm$0.7		&	2.83	& 7.7 	\\
NW2     		&    	2.31   	   &   	0.36  		&    	1.34    	& 0.8$\pm$0.4		&	2.81	& 5.2 	\\
NW3    		&     	0.38  	   &  	--  			&     	1.32   	& 1$\pm$0.6		&    	--   	& 4.4		\\
NW4    		&     	1.17  	   &   	-- 			&     	1.23   	& 2$\pm$0.7		&    	--   	& 3.4		\\
NW5   		&     	2.71  	   &	0.42  		&     	1.21   	& 2.2$\pm$0.5		&    	2.86  & 4.7	 \\
SE1      		&    	1.20   	   &   	1.46   		&    	1.42    	& 0.2$\pm$0.1		&    	2.87	 & 6.7    	\\
\hline             
\end{tabular}
\label{flux-ratios}
\end{table}

\begin{table}
\center
\caption{Projected lengths of known atomic jets in very low-mass stars and brown dwarfs}
\begin{tabular}{lllllll}
\hline
\hline
Object 				&   	RA (J2000)	&   	Dec (J2000)  	& 	Class 	&	Length (pc) & Ref\tablenotemark{a} \\
\hline
M1701117 (HH 1165)    	&	05:40:25.8     	&	-02:48:55.4 	& 	I 		&	0.26	& [1]	\\ 
M1082188 (HH 1158)   	&    	05:38:34.5 	& 	-02:53:51.5 	&	Flat 		&	0.005	&  [2]	\\
ISO 143   				&	11:08:22.4 	&	-77:30:27.7 	& 	II 		&	$<$0.001	&  [3] \\
ISO ChaI 217      		& 	11:09:52.2 	&	-76 39 12.81 	&	II 		&	0.002	&  [4]	\\
2MASS1207A    		&  	12:07:33.4 	& 	-39:32:54.0  	&	II 		&	0.004	&  [5]	\\
Par-Lup3-4 (HH 660)     	& 	16:08:51.4 	& 	-39:05:30.5 	&	II 		&	0.004	&  [6]	\\ 
ISO-Oph 32     			& 	16:26:22.0 	& 	-24:44:38 		& 	II 		&	$<$0.001	&  [4]	\\
ISO-Oph 102     		& 	16:27:06.6 	& 	-24:41:48.8  	&	II 		&	$<$0.001	&  [7]	\\
LS-RCr A1 			&	19:01:33.7  	&	-37:00:30.0 	&	II 		&	$<$0.001	&  [8]	\\
\hline             
\end{tabular}
\tablenotetext{a}{References for the length are: [1] This work; [2] Riaz \& Whelan 2015; [3] Joergens et al. (2012); [4] Whelan et al. (2009a); [5] Whelan et al. (2012); [6] Whelan et al. (2014); [7] Whelan et al. (2005); [8] Whelan et al. (2009b).  }
\label{known-jets}
\end{table}

%---------------------------------------------------

\clearpage

\counterwithin{figure}{section}

\section{Figures}

\begin{figure}[ht]
\center
\includegraphics[scale=0.45]{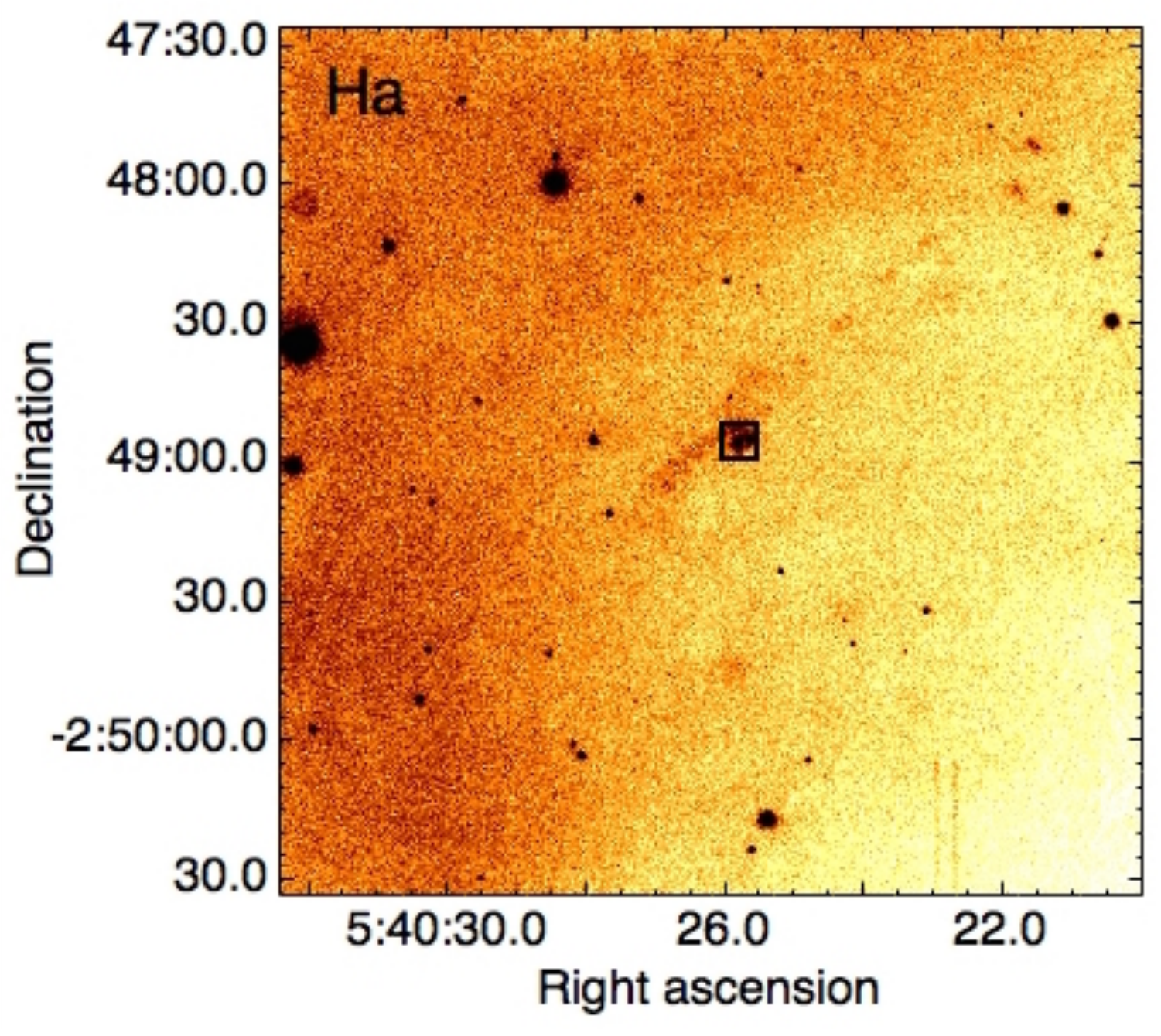}  \\  \vspace{0.2in}
\includegraphics[scale=0.45]{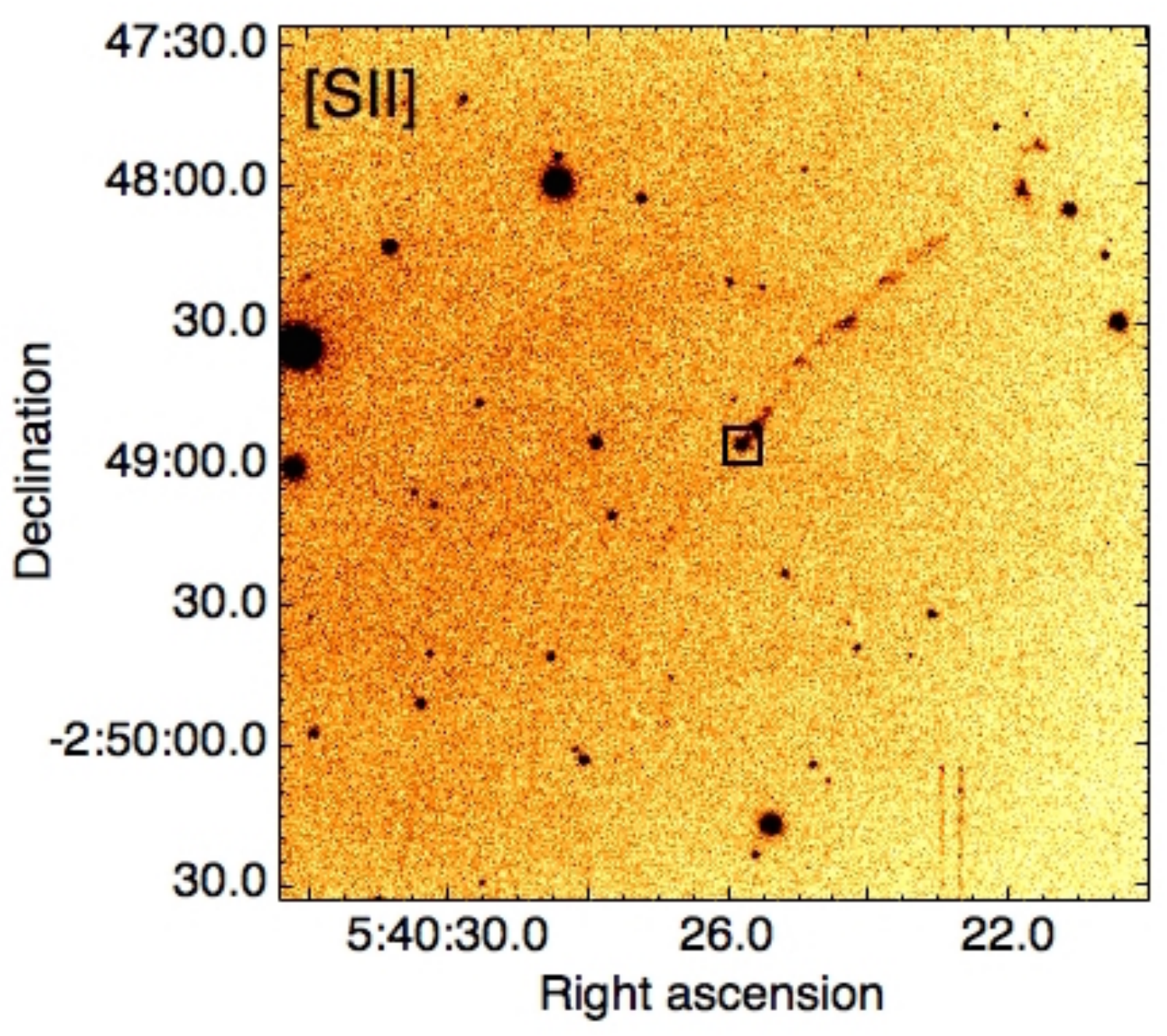}
\caption{HH~1165 jet in the H$\alpha$ (top) and [SII] (bottom) images. These are color-inverted images where the darker shades represent the denser region. The image size is 3$\arcmin \times$3$\arcmin$ ($\sim$0.34x0.34 pc). North is up, east is to the left. The driving source M1701117 is marked by a black box point.  }
\label{Ha-SII-img}
\end{figure}

\begin{figure}[h]
\center
\includegraphics[scale=0.4]{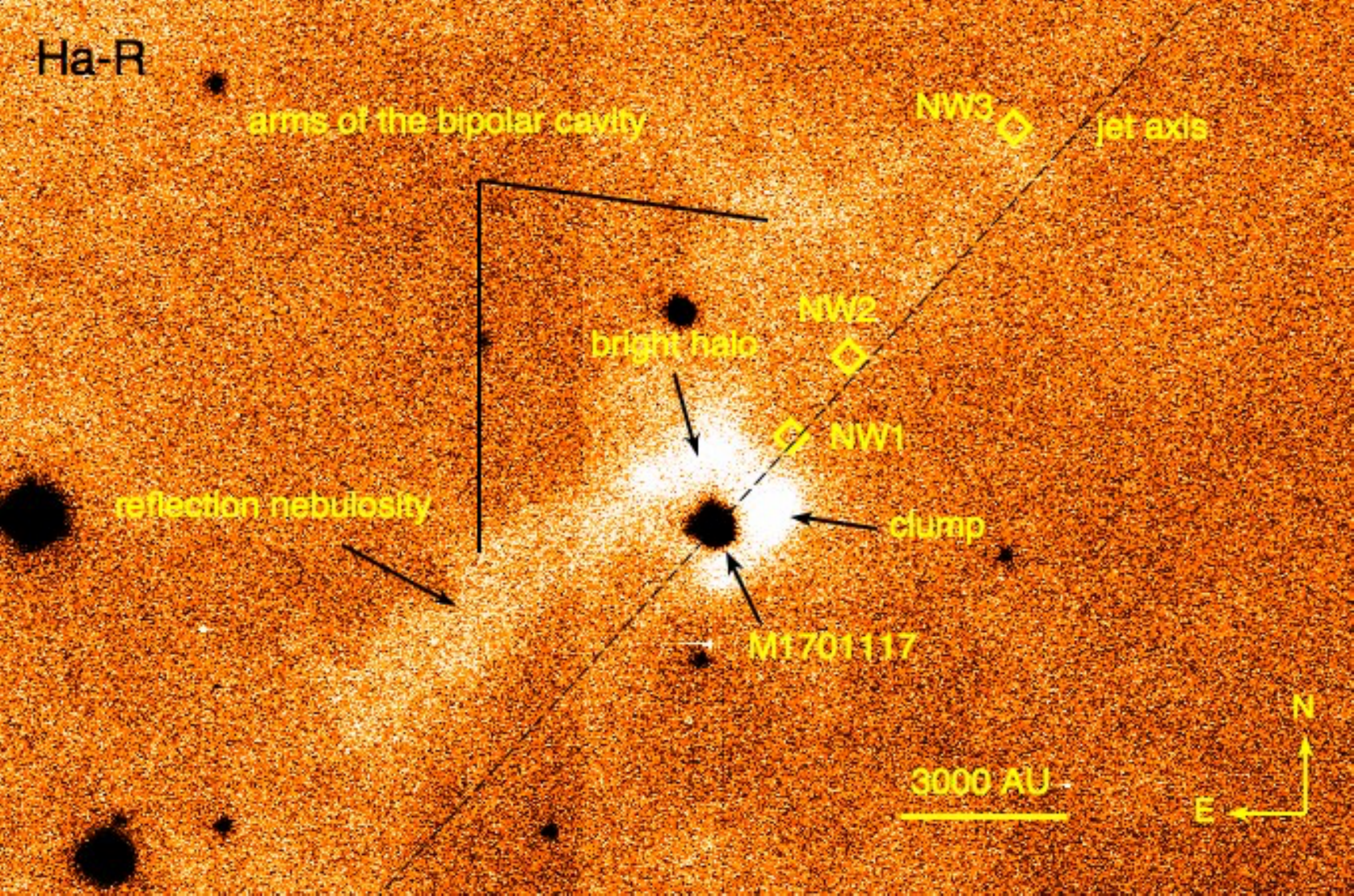}  \\ \vspace{0.3in}
\includegraphics[scale=0.4]{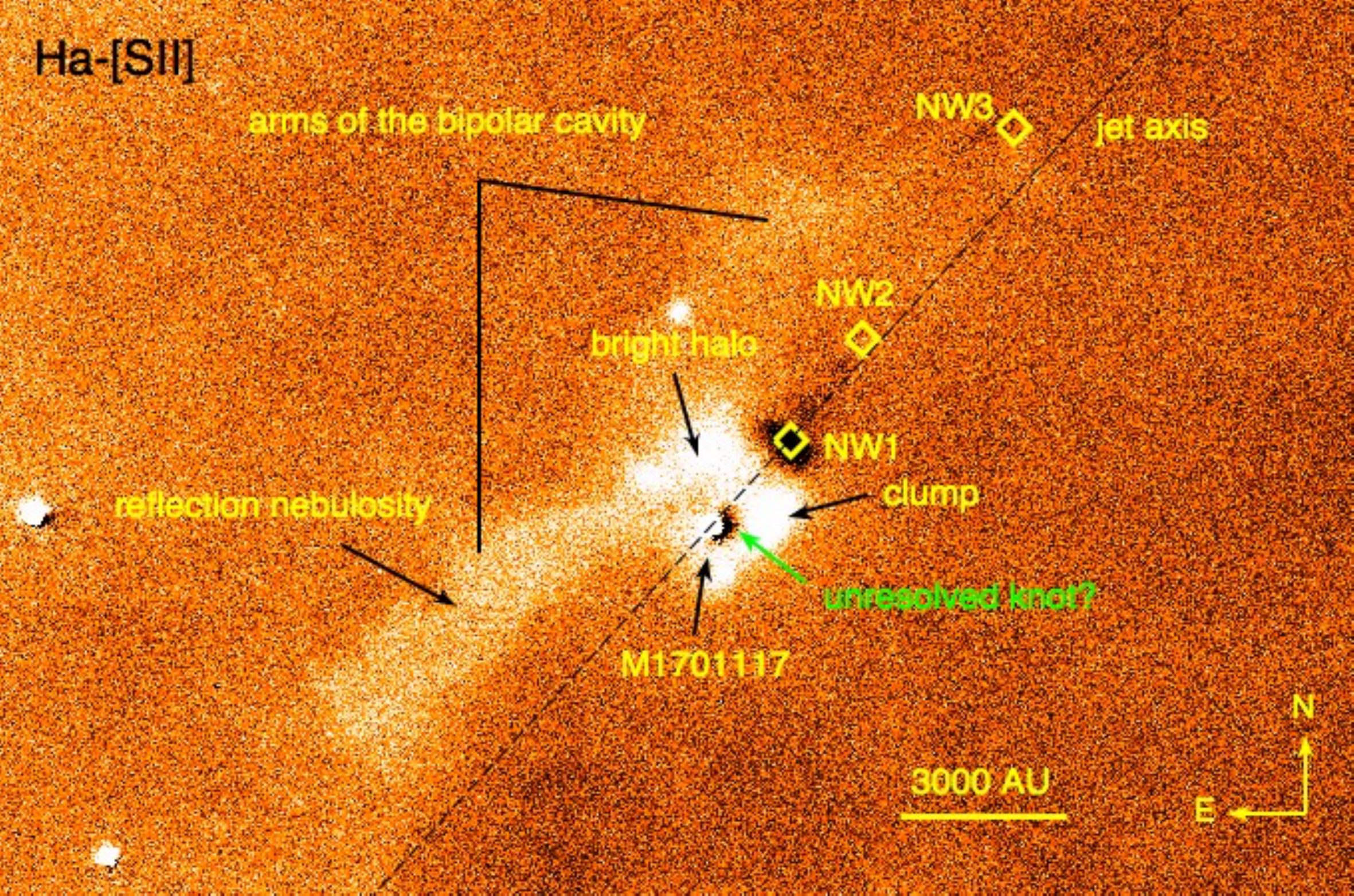}
\caption{The (H$\alpha$-$R$) continuum-subtracted image (top) and the (H$\alpha$-[SII]) image (bottom) for HH 1165. The labels and image size are the same as in Fig.~\ref{Ha-img}. We have roughly marked the bright reflection nebulosity tracing the outflow cavity on the eastern side of the source, the bright halo surrounding the driving source, and the bright clump seen within the halo. Solid line is the jet axis close to the driving source (PA$\sim$316$^{\circ}$). Also marked are the locations of the inner most NW1-3 knots as seen in the [SII] image. This is a zoomed-in 1$\arcmin \times$1$\arcmin$ image centered on M1701117. The linear scale is shown at the bottom. North is up, east is to the left. }
\label{R-Ha-img}
\end{figure}

\begin{figure}
\centering
\includegraphics[width=5cm]{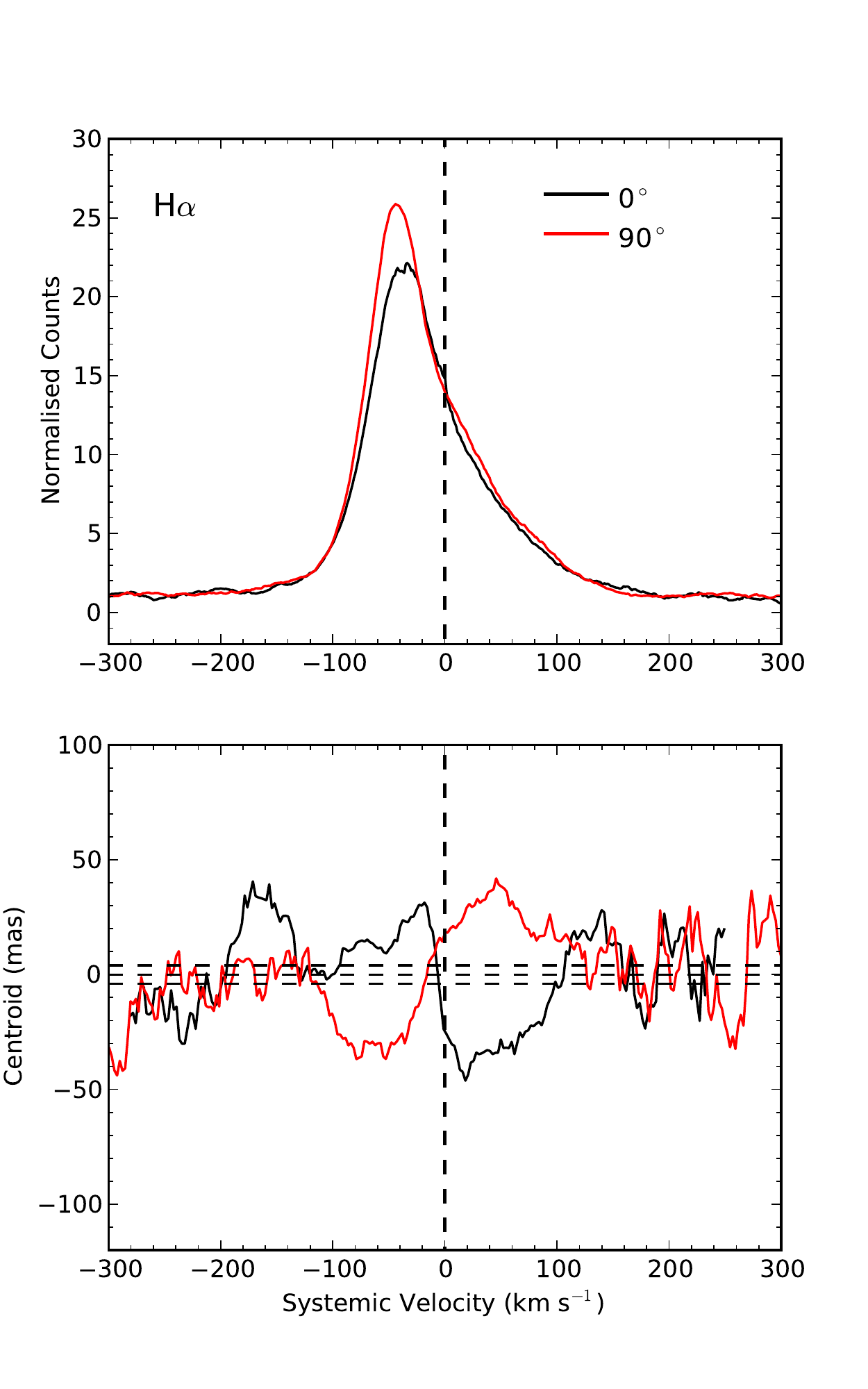}
\includegraphics[width=5cm]{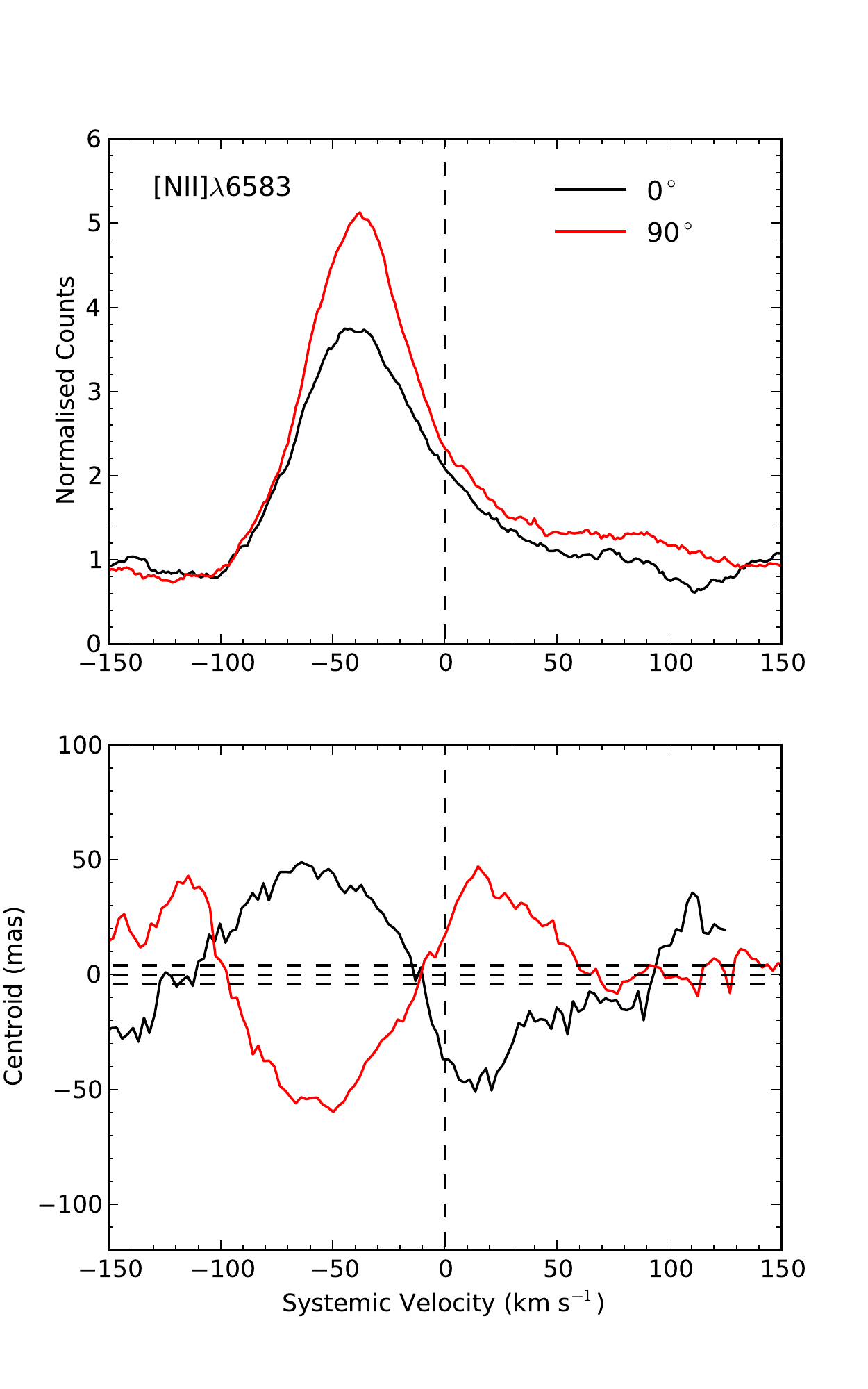}
\includegraphics[width=5cm]{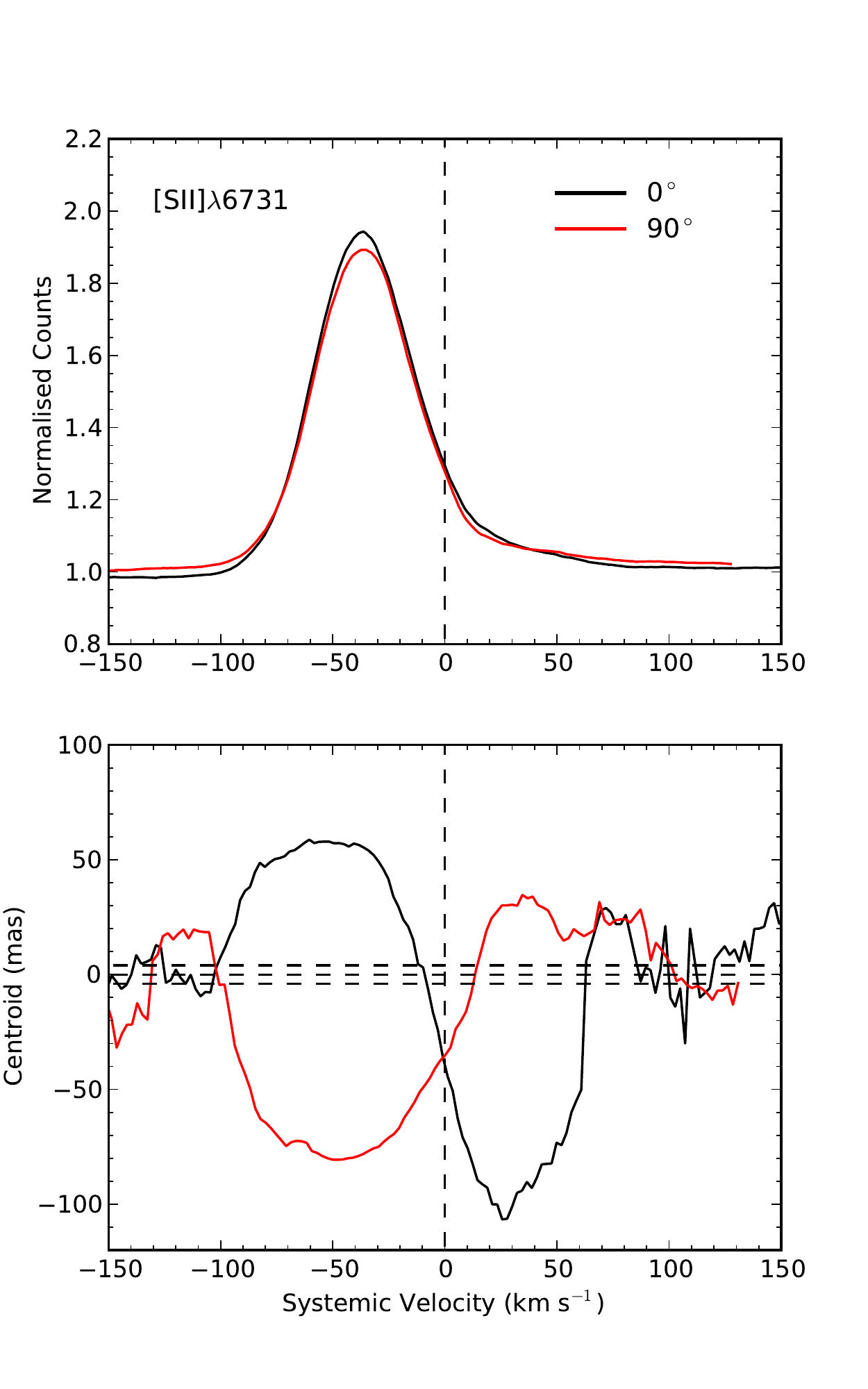}
\caption{Spectro-astrometric analysis of the H$\alpha$, [NII]$\lambda$6583 and [SII]$\lambda$6731 lines. The red lines are the east-west spectra (90$^{\circ}$ slit PA), and the black-lines are the north-south spectra (0$^{\circ}$ slit PA). The spectra have been binned to increase the SNR. The horizontal dashed lines delineate the $\pm$ 1-$\sigma$ uncertainty in the peak of the line emission. The uncertainty in the continuum is much higher and reaches $\sim$ 20 mas. }
\label{spectro}
\end{figure}

\begin{figure*}[h]
\centering
\includegraphics[scale=0.4,angle=-90]{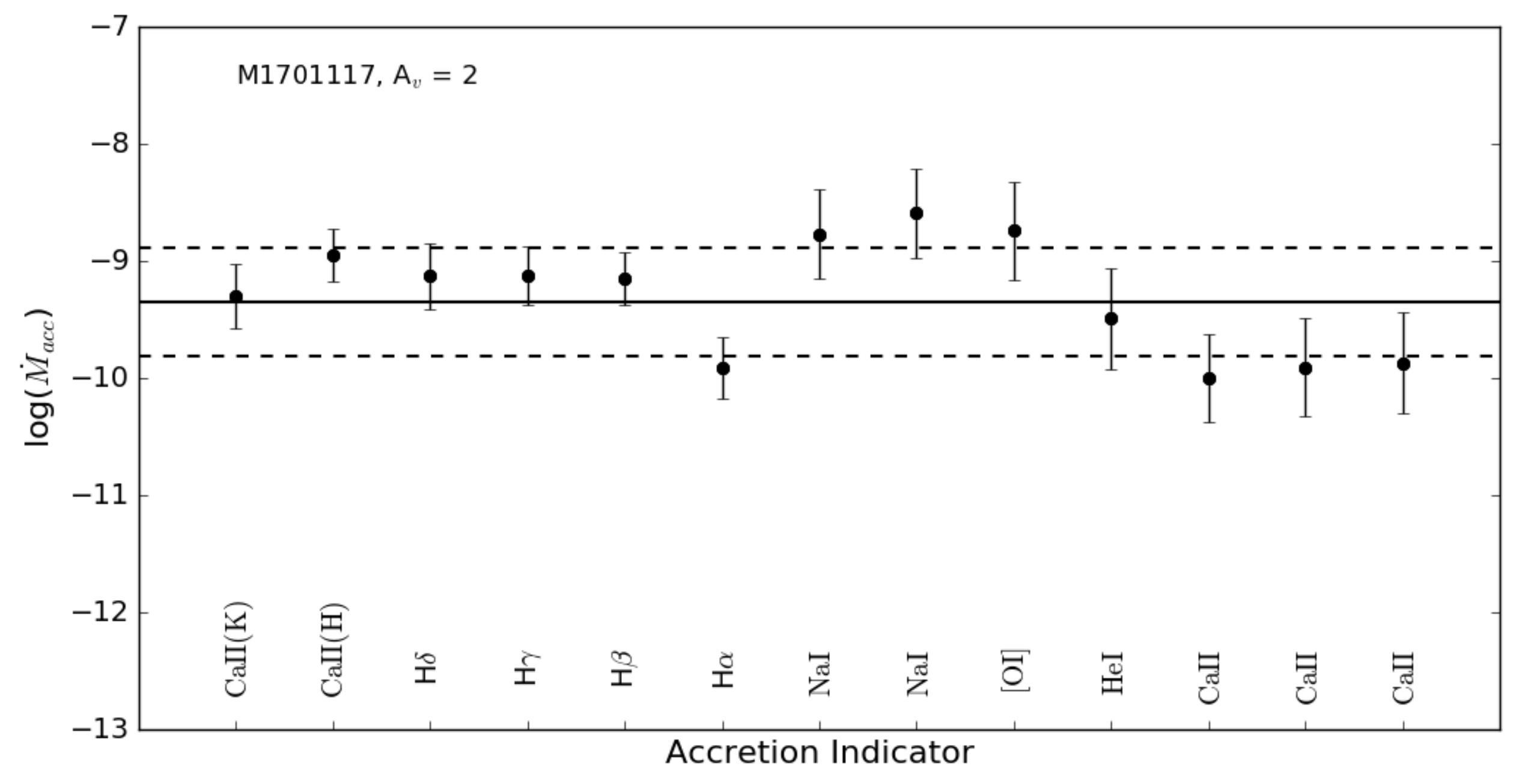}
\caption{The accretion rates $\log$ $\dot{M}_{acc}$ in $M_{\sun}$ yr$^{-1}$ derived from various line diagnostics (labelled) for M1701117. Horizontal line indicates the mean level calculated without including the [O~{\sc i}] line. }
\label{accretion}
\end{figure*}

\begin{figure}[h]
\centering
\includegraphics[scale=0.6]{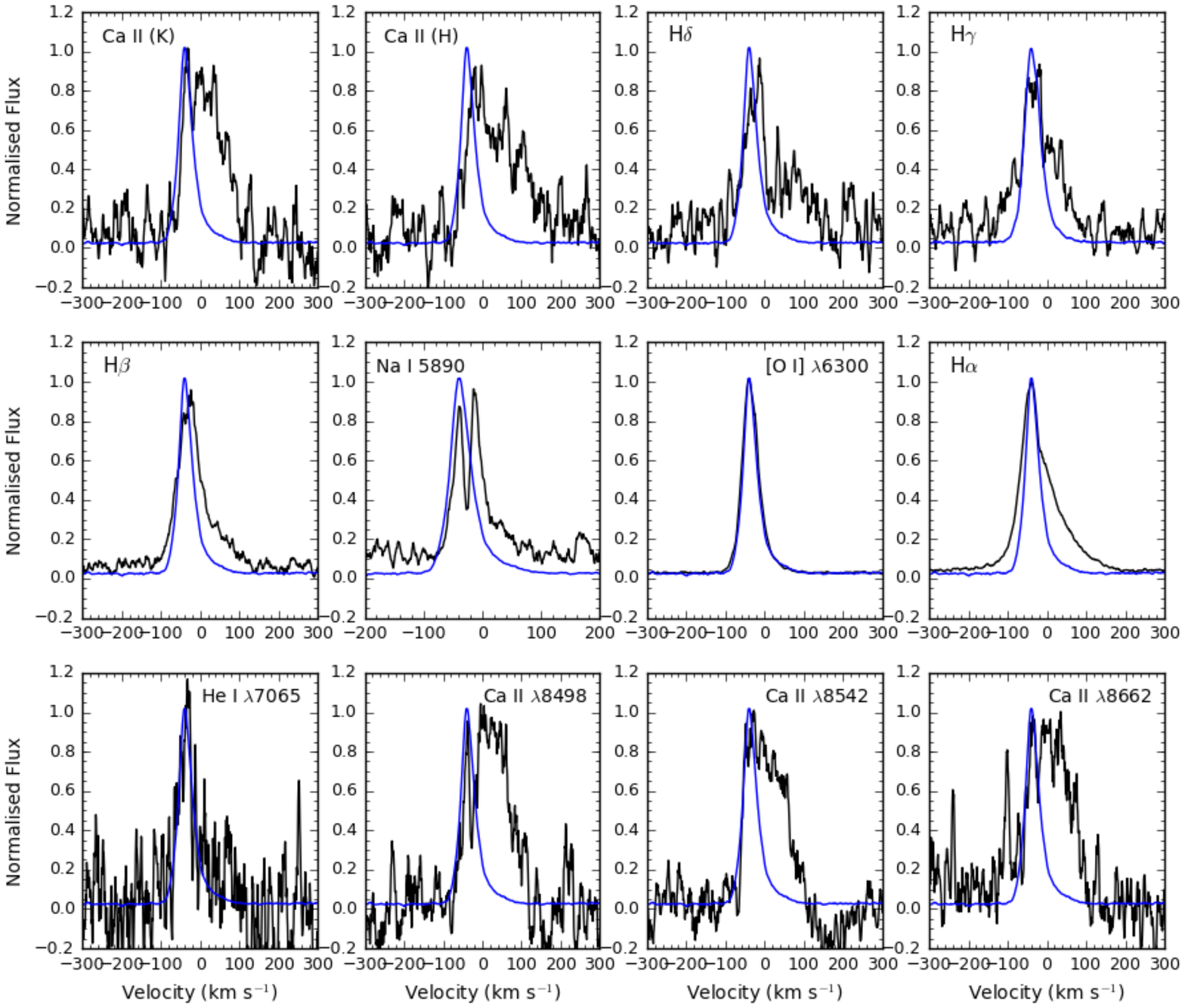}
\caption{A comparison of the shape of all accretion line diagnostics (black) used to calculate $\dot{M}_{acc}$ with the shape of the outflow-associated [SII]$\lambda$ 6731 line (blue).  }
\label{accretion_lines}
\end{figure}

\begin{figure}
\centering
\includegraphics[scale=0.55]{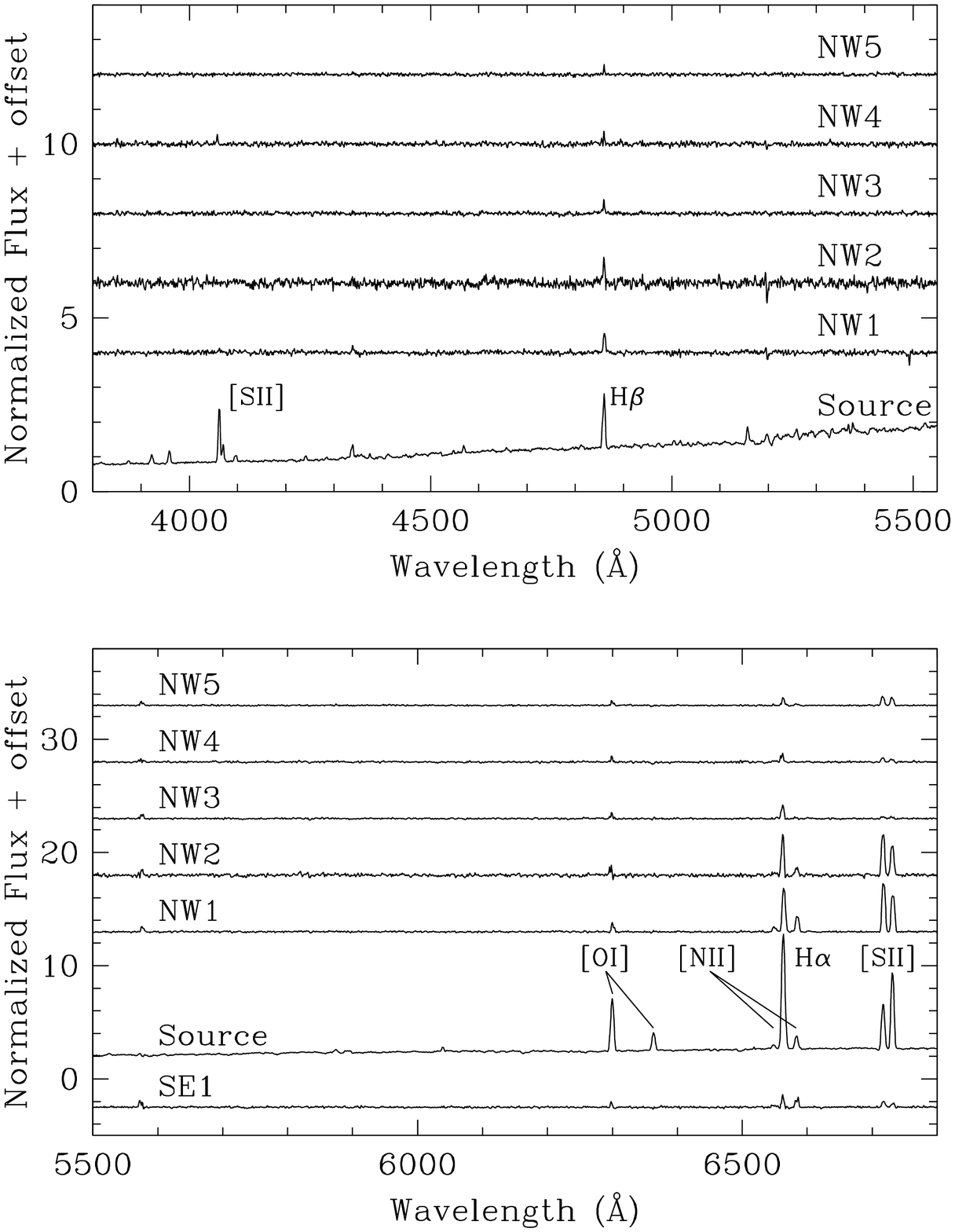}
\caption{The SOAR/Goodman HTS sky-subtracted spectra of the M1701117 driving source and the various knots along the HH1165 jet. For the SE1 feature we only show the spectrum redward of 5500 {\AA} since it is featureless at shorter wavelengths. }
\label{goodman}
\end{figure}

\begin{figure}
\center
\includegraphics[scale=0.6]{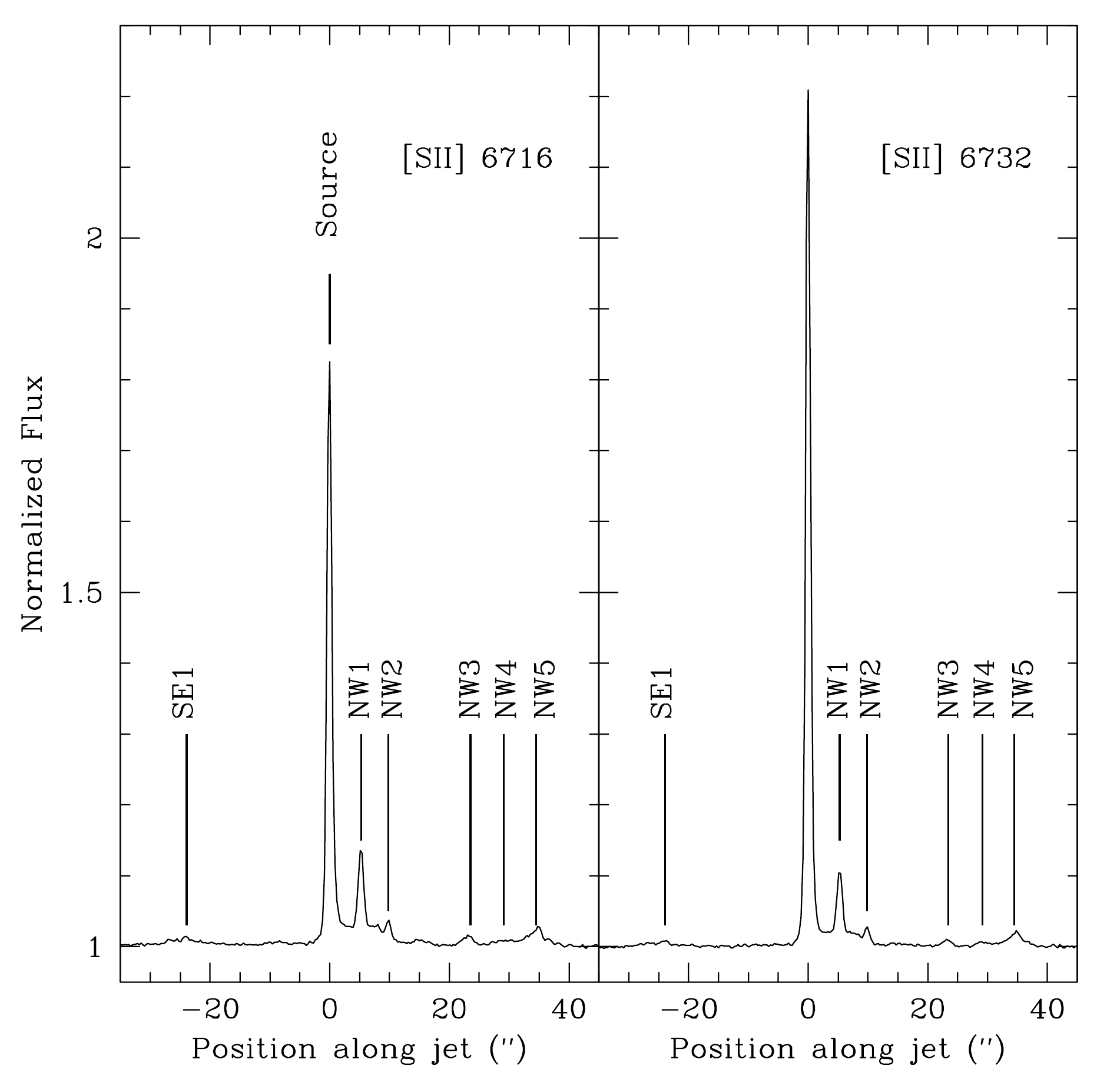}
\caption{The [SII]$\lambda\lambda$6716, 6731 relative line strength and width at the source position and various knots along the jet. }
\label{strength-SII}
\end{figure}

\begin{figure}
\center
\includegraphics[scale=0.6]{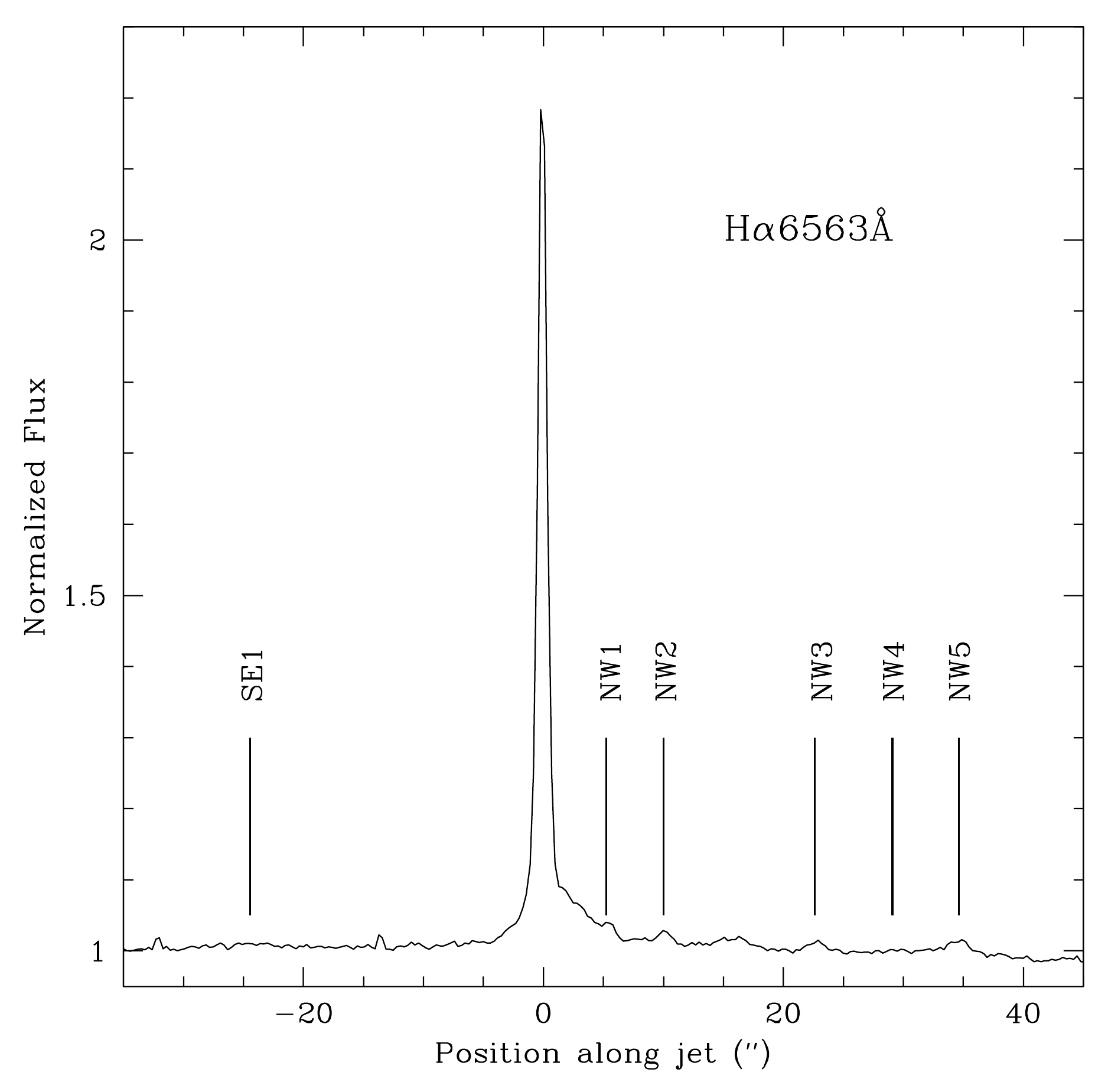}
\caption{The H$\alpha$ relative line strength and width at the source position and various knots along the jet. }
\label{strength-Ha}
\end{figure}

\begin{figure}
\centering
\includegraphics[scale=0.5]{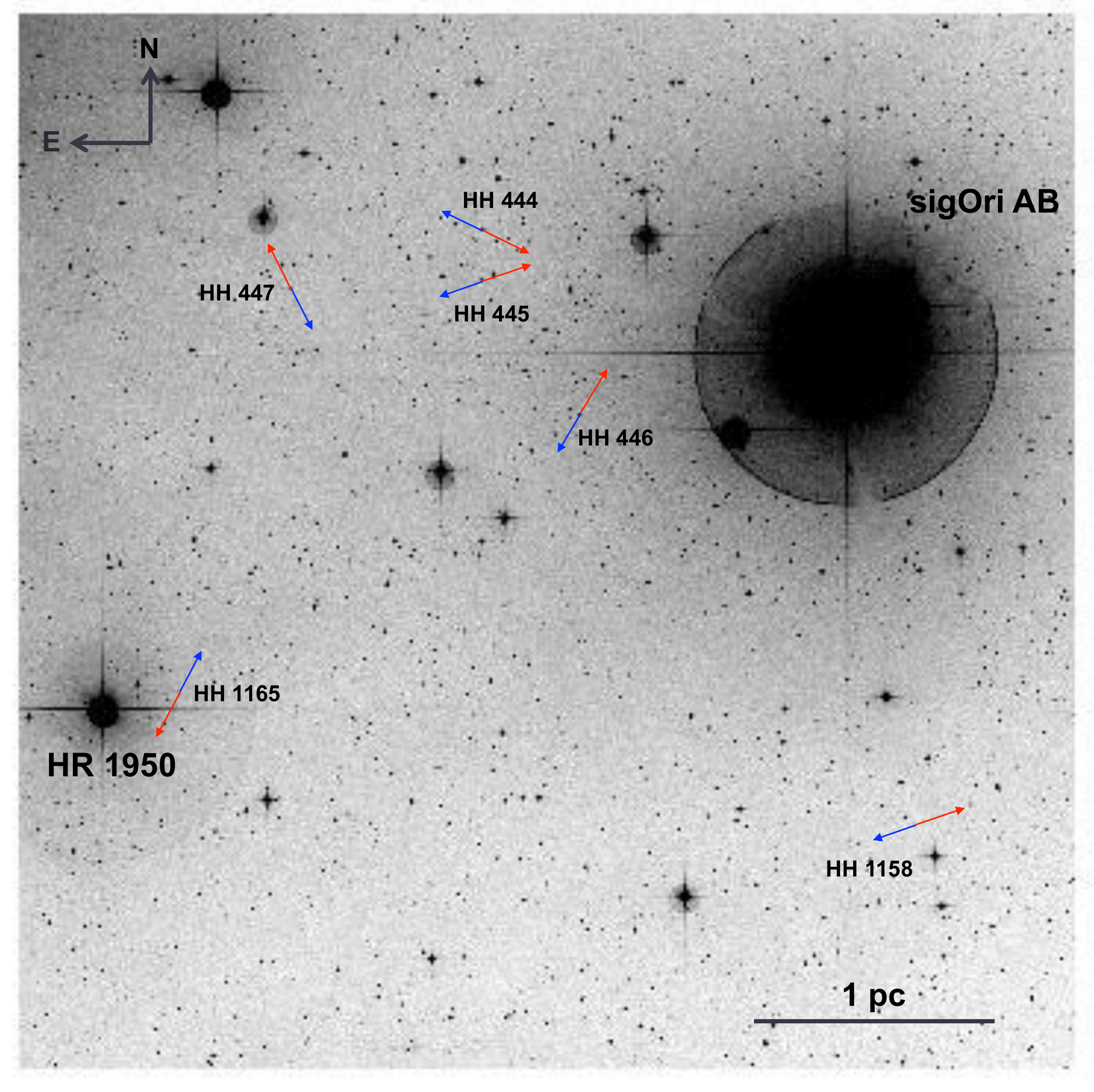}
\caption{The location of HH 1165 jet with respect to the $\sigma$ Orionis core and HR~1950. Also shown are the previously known HH jets in the $\sigma$ Orionis cluster. The blue and red arrows indicate the PA of the blue- and red-shifted lobes of the jets, respectively. North is up, east is to the left. The image size is $\sim$40$\arcmin \times$40$\arcmin$.}
\label{position}
\end{figure}

\end{document}